\documentclass[a4paper,11pt]{article}
\pdfoutput=1 

\usepackage{jheppub} 
\usepackage{amsmath,amssymb,amsthm,amscd,graphicx}
\usepackage{hyperref}
\usepackage{suffix}
\usepackage{mathtools}
\input epsf.sty

\newtheorem{theorem}{Theorem}[section]

\theoremstyle{definition}

\newtheorem{remark}[theorem]{Remark}

\newcommand{\CA}{{\cal A}}

\newcommand{\CC}{{\cal C}}

\newcommand{\CF}{{\cal F}}
\newcommand{\CG}{{\cal G}}

\newcommand{\CN}{{\cal N}}
\newcommand{\CO}{{\cal O}}

\newcommand{\CS}{{\cal S}}

\def\IZ{{\mathbb Z}}
\def\IR{{\mathbb R}}
\def\IC{{\mathbb C}}

\def\IP{{\mathbb P}}

\def\IS{{\mathbb S}}
\def\IF{{\mathbb F}}

\newcommand{\re}{{\rm e}}
\newcommand{\ri}{{\rm i}}
\newcommand{\rd}{{\rm d}}

\newcommand{\bx}{\boldsymbol{x}}
\newcommand{\mZ}{\mathsf{Z}}
\newcommand{\mF}{\mathsf{F}}
\newcommand{\mfF}{\mathfrak{F}}

\newcommand{\mD}{\mathfrak{D}}

\newcommand{\be}{\begin{equation}}
\newcommand{\ee}{\end{equation}}
\newcommand{\ba}{\begin{aligned}}
\newcommand{\ea}{\end{aligned}}
\newcommand{\ben}{\begin{eqnarray}\displaystyle}
\newcommand{\een}{\end{eqnarray}}

\newcommand{\sectiono}[1]{\section{#1}\setcounter{equation}{0}}

\newdimen\tableauside\tableauside=1.0ex
\newdimen\tableaurule\tableaurule=0.4pt
\newdimen\tableaustep
\def\phantomhrule#1{\hbox{\vbox to0pt{\hrule height\tableaurule width#1\vss}}}
\def\phantomvrule#1{\vbox{\hbox to0pt{\vrule width\tableaurule height#1\hss}}}
\def\sqr{\vbox{%
  \phantomhrule\tableaustep
  \hbox{\phantomvrule\tableaustep\kern\tableaustep\phantomvrule\tableaustep}%
  \hbox{\vbox{\phantomhrule\tableauside}\kern-\tableaurule}}}
\def\squares#1{\hbox{\count0=#1\noindent\loop\sqr
  \advance\count0 by-1 \ifnum\count0>0\repeat}}
\def\tableau#1{\vcenter{\offinterlineskip
  \tableaustep=\tableauside\advance\tableaustep by-\tableaurule
  \kern\normallineskip\hbox
    {\kern\normallineskip\vbox
      {\gettableau#1 0 }%
     \kern\normallineskip\kern\tableaurule}%
  \kern\normallineskip\kern\tableaurule}}
\def\gettableau#1{\ifnum#1=0\let\next=\null\else
\squares{#1}\let\next=\gettableau\fi\next}

\DeclarePairedDelimiterX\MeijerM[3]{\lparen}{\rparen}%
{\begin{smallmatrix}#1 \\ #2\end{smallmatrix}\delimsize\vert\,#3}

\newcommand\MeijerG[8][]{%
  G^{\,#2,#3}_{#4,#5}\MeijerM[#1]{#6}{#7}{#8}}

\WithSuffix\newcommand\MeijerG*[7]{%
  G^{\,#1,#2}_{#3,#4}\MeijerM*{#5}{#6}{#7}}

\newcommand{\figref}[1]{Fig.~\protect\ref{#1}}
\title{\Huge{\boldmath Large $N$ instantons, BPS states, and the replica limit}}

\author{Marcos Mari\~no and Maximilian Schwick}

\affiliation{D\'epartement de Physique Th\'eorique et Section de Math\'ematiques\\
Universit\'e de Gen\`eve, Gen\`eve, CH-1211 Switzerland}

\abstract{We study the relation between large $N$ instantons and conventional instantons, focusing on   
matrix models and topological strings. We show that the resurgent properties of the perturbative series at fixed 
but arbitrary $N$, including the replica limit $N \rightarrow 0$, can be obtained from large $N$ instantons. In the case of topological strings, 
it has been conjectured that 
the resurgent structure encoded by large $N$ instantons is closely related to 
the spectrum of BPS states. We give direct evidence 
for this connection in the case of Seiberg--Witten theory and other topological string 
models, and we show in detail how the resurgent 
properties at fixed $N$ follow from the large $N$ theory, and 
therefore can be used to obtain information on BPS invariants.}    

\begin{document}
\maketitle
\flushbottom

\sectiono{Introduction}

In many quantum theories, the number of constituents $N$ can be regarded as a 
parameter. It has been known since 1970 that the limit in which $N$ is large might lead to insights which are 
not easily available otherwise. Physical observables can be expanded systematically in a power 
series in $1/N$, and this expansion is an 
alternative to the perturbative expansion in the coupling 
constant. Both expansions share however many important 
formal aspects. For example, the $1/N$ expansion is 
generically asymptotic, and one expects to have non-perturbative effects which are exponentially small in $N$. 
A known source for these corrections are large $N$ instantons (see e.g. the last chapter of \cite{mmbook} for an exposition). 
In some simple examples, like matrix models, large $N$ instantons can be described in detail, and they control the large order 
behavior of the $1/N$ expansion. 

Although $N$ is the number of constituents, 
and is therefore a positive integer, in the conventional perturbative series in the coupling constant one can take $N$ to 
be an arbitrary complex number, since the coefficients of this series are typically polynomials in $N$. 
 ``Unphysical" values of $N$ turn out to be useful. For example, the case $N=0$, 
which we will call in this paper the {\it replica limit}, can be used to 
study percolation or quenched disordered systems. It is natural to 
ask how instanton effects associated to the asymptotic behavior of the perturbative 
series depend on $N$, for arbitrary complex values. Since many instanton 
calculations depend in principle on the fact that $N$ is a positive integer, extrapolation to other values might be subtle, as illustrated e.g. in \cite{mckane,bray}. 

In this paper we will address some of these questions. Our leitmotiv will be the relationship 
between the instantons associated to the $1/N$ expansion, and the 
instantons associated to the perturbative series in the coupling constant. We will refer to the former as large $N$ instantons, 
and to the latter as {\it conventional} instantons, or instantons of the perturbative series. We will obtain 
results for conventional instantons from large $N$ instanton amplitudes. As we will see, the latter implement automatically 
the required analytic continuation to arbitrary 
complex $N$, provided one regularizes the amplitudes in an appropriate way. In particular, we will show that, 
somewhat amusingly, large $N$ instantons make it possible to calculate instanton amplitudes in the replica limit at $N=0$. 
From the point of view of the theory of resurgence, we are relating the so-called ``parametric" resurgent properties 
of the $1/N$ expansion, to the conventional resurgent 
properties of the perturbative series. In physics, a similar situation arises when one compares the 
resurgent structure of WKB expansions to the resurgent structure of conventional perturbation theory (see \cite{gm-peacock} for a 
definition of resurgent structure, and \cite{mmbook,ss,msauzin, abs} for surveys of the theory of resurgence). 

The examples we will focus on are Hermitian matrix models and topological string theories, 
as well as some close cousins thereof. One of the reasons to look at these examples is 
simply that their large $N$ instantons are now known in detail. A deeper reason is the following. It has been 
conjectured that the ``parametric" resurgent structure of topological 
strings on Calabi--Yau (CY) threefolds is closely related to the spectrum of BPS states 
and to BPS invariants (like Donaldson--Thomas invariants) of the CY. 
The calculation of the parametric resurgent properties is difficult, and one could 
think that conventional perturbative 
series are more amenable to an analytic treatment, and therefore could be used to 
determine e.g. BPS invariants. This was the motivation 
for the paper \cite{gm-peacock}, which studied perturbative series in the string coupling 
constant obtained from the $1/N$ expansion or the genus expansion of 
the topological string free energy. In particular, \cite{gm-peacock} calculated the conventional 
instanton amplitudes for the perturbative series, and in particular its associated Stokes constants, in many examples. 
However, in order to make contact with the original, parametric resurgent structure of the topological string, one needs to 
clarify the relation between the perturbative series and their large $N$ counterparts. 
This is one of the results we will obtain in this paper. In particular, we explain many of the findings in \cite{gm-peacock} from first principles. We find that the integer Stokes constants obtained in \cite{gm-peacock} are given by weighted sums 
of the Stokes constants of the full-fledged topological string theory, which are conjecturally given by BPS invariants. 
The sum is over all possible charges of the D4 brane which becomes massless at the conifold point. 
 
Although the relation between parametric and conventional resurgence 
reduces in some simple cases to a change of variables, in the physical applications
 considered here one faces more subtle situations which have not been addressed in the 
 mathematical literature. First of all, 
 large $N$ series and large $N$ instanton amplitudes turn out to be  
 singular when one wants to make contact with conventional 
 perturbative results. This is essentially due to the conifold-like behavior 
 of the amplitudes. As mentioned above, this requires 
 a regularization procedure which was 
 already encountered in \cite{multi-multi}. Although there is a natural way to 
 implement such a regularization, this is not a completely straightforward  
 step and needs further mathematical justification. In this paper we will 
 simply show that our regularization leads to results in agreement with the empirical data. 
 A more delicate issue is the following. In parametric resurgence one can 
 have wall-crossing phenomena, namely, the resurgent structure can 
 change discontinuously as we move in moduli space. This was noted already in \cite{reshyper}, 
 anticipating further developments in \cite{ks,gmn}. 
 It turns out that, in most examples, the perturbative limit corresponds to a point in the 
 curve of marginal stability, i.e. to a point where precisely the 
 resurgent structure changes discontinuously. The resurgent structure of the perturbative 
 series seems to pick one of the resurgent structures appearing close to the curve of marginal stability, and with our current understanding we can not determine {\it a priori} which one. We hope that our results will stimulate rigorous research on these issues.

This paper is organized as follows. In section \ref{sec-warmup} we consider two 
simple models where the relation between 
large $N$ instantons and conventional instantons can be worked out in detail. The first one 
is an $O(N)$ vector model in zero dimensions considered previously by Br\'ezin and Hikami 
\cite{hikami-brezin}, and the second one is Chern--Simons theory on the three-sphere, 
where the importance of appealing to large $N$ instantons will be, hopefully, clear. In section \ref{sec-gt} we present the 
general relationship between large $N$ instantons and conventional instantons in matrix 
models and in topological strings on local CY manifolds, based on the results in \cite{multi-multi} 
and the more recent ones in \cite{gm-multi,im}. The rest of the paper is devoted to examples 
illustrating the general theory. Section \ref{sec-mm} considers the case of the one- and two-cut cubic 
matrix model. Section \ref{sec-sw} studies the topological string associated to the Seiberg--Witten (SW) 
curve. We first give new direct evidence in this model for the conjecture that the resurgent structure 
of the topological string contains information about the spectrum of BPS states, and then 
we consider its relation to the resurgent structure of the perturbative series. Section \ref{sec-cy} studies examples of local 
CY threefolds. In particular, we revisit the examples of \cite{gm-peacock} and use the results of section \ref{sec-gt} 
to clarify the relationship between the Stokes constants obtained in that paper and the conjectural BPS invariants of the CY threefold. 
A final section contains some conclusions and prospects for the future. We have added two short Appendices. The first one outlines a simple 
derivation of the multi-instanton formulae of \cite{gm-multi,im}, by extending the approach of \cite{abk} to trans-series. The second one shows how to derive the multi-instanton formula for the one-cut matrix model given in \cite{multi-multi} from the general results obtained with topological string techniques in \cite{gm-multi, im}.

\begin{remark} In this paper we will consider three types of quantities and 
two different types of formal power series. On one hand, we will consider  
the standard $1/N$ expansion, whose coefficients are functions of an appropriate 't Hooft parameter, and we will also consider conventional perturbative series, whose coefficients are functions of $N$. We will use $\CF$, $Z$ for the free energies and partition functions of the former, and the notation $\mathfrak{F}$, $\mathfrak{Z}$ for the latter. 
On the other hand, exact, non-perturbative free energies or partition functions will be 
denoted by $\mathsf{F}$ or $\mathsf{Z}$, respectively. 
\end{remark}

\sectiono{Warm-up examples}
\label{sec-warmup}

\subsection{A toy model with $O(N)$ symmetry}

Our first example is very simple but hopefully it will illustrate some of the issues at hand. It is also a solvable example in the sense that 
all results can be obtained in a completely straightforward way and from first principles. This example is the zero-dimensional version of the $\varphi^4$ model with an $O(N)$ symmetry, which was considered in e.g. \cite{hikami-brezin} from the point of view of instanton physics (see also \cite{klebanov} for a nice survey of the large $N$ 
limit of this model). It is defined by the partition function 
\be
\label{zng}
\mathsf{Z} (g;N)= {1\over (2 \pi)^{N/2}} \int_{\IR^N} \rd \bx \, \re^{-S(\bx)}, \qquad S(\bx)= {\bx^2 \over 2} + {g\over 4} (\bx^2)^2. 
\ee
This function has an asymptotic expansion for small $g$ and fixed $N$ given by 
\be
\label{phiN}
\mathfrak{Z}(g;N) = \sum_{k \ge 0} a_k(N) g^k, \qquad a_k(N)= (-1)^k  {\Gamma(2k+N/2) \over k! \Gamma(N/2)}. 
\ee
The free energy of the theory is defined by
\be
\mathsf{F}(g;N)=\log \mathsf{Z}(g;N),  
\ee
and it leads to a formal perturbative series 
\be\label{pfe-qvm}
\mathsf{F}(g,N)\sim \mathfrak{F}(g;N)=  \sum_{k \ge 1} a^{(c)}_k (N) g^k. 
\ee
This theory has a ``replica limit" $N \to 0$. It is characterized by the perturbative series 
\be
\label{quartic-replica}
\mathfrak{R}(g)= \lim_{N \to 0} { \mathfrak{Z}(g;N)-1 \over N}=\lim_{N \to 0} {  \mathfrak{F}(g;N) \over N}= \sum_{k \ge 1} {(-1)^k (2k-1)!  \over 2 \, k!} g^k. 
\ee
At the same time, one can consider the standard large $N$ limit, in which $N$ is large but the 't Hooft parameter 
\be
t= g N
\ee
remains fixed. In this regime, the free energy has the asymptotic $1/N$ expansion 
\be
\label{F1N}
{1\over N} \mathsf{F} \sim \CF(t;N)= \sum_{k \ge 0} \CF_k (t) N^{-k}. 
\ee
The functions $\CF_k (t)$ can be computed, as a power series in $t$, directly from the perturbative series. 
One finds, at the very first orders, 
\be
\label{small-lam}
\ba
\CF_0(t)&= -{t \over 4}+ {t^2 \over 4} - {5 t^3 \over 12}+{7 t^4 \over 8}+ \CO(t^5), \\
\CF_1 (t)&=-{t \over 2} + {5 t^2 \over 4} -{11 t^3 \over 3} +{93 t^4 \over 8}+  \CO(t^5), \\
\CF_2(t)&={3 t^2 \over 2}-{32 t^3 \over 3} +{117 t^4 \over 2}+   \CO(t^5). 
\ea
\ee
They can be also computed in closed form, as functions of $t$, by using standard tricks. We will follow here the procedure in 
\cite{hikami-brezin}. Since the partition function has an $O(N)$ symmetry, we first go to spherical coordinates to write
\be
\mZ(g;N)= {{\rm vol} (\IS^{N-1}) \over (2 \pi)^{N/2}} \int_0^\infty \rd r \, r^{N-1} \re^{-{1\over 2} r^2 -{g \over 4} r^4}. 
\ee
We now change variables to 
\be
r= N^{1/2} \re^{\xi/2}, 
\ee
and we find
\be
\label{zlargeN}
\mZ(g;N)= {(N/2)^{N/2} \over \Gamma(N/2)} \int_\IR \re^{-N f(\xi)} \, \rd \xi, 
\ee
where
\be
f(\xi)= {1 \over 2}\re^\xi +{t \over 4} \re^{2\xi} -{\xi\over2}. 
\ee
One advantage of this formulation is that the derivatives of $f(\xi)$ can be computed in closed form:
\be
f^{(n)}(\xi)= {1 \over 2}\re^\xi+2^{n-2} t \re^{2\xi} -{1\over2} \delta_{n1}, \qquad n \ge 1. 
\ee
We can now apply to (\ref{zlargeN}) a standard saddle-point 
evaluation for $N$ large, in order to obtain the $1/N$ expansion. The critical point 
$\xi_c$ which is relevant to reproduce the 
perturbative expansion is
\be
\xi_c =\log\left[ {(1+ 4 t)^{1/2}-1 \over 2 t} \right]. 
\ee
We will denote
\be
f^{(n)}_c\equiv f^{(n)}(\xi_c), 
\ee
and one finds
\be
\ba
\CF(t;N)&\sim{1\over N} \log\left[ { {(N/2)^{N/2} \over \Gamma(N/2)}} {\sqrt{2 \pi \over N}} \right]
 - f_c -{1\over 2 N} \log(f''_c) + {1\over N^2} \left( {5 \over 24} {(f^{(3)}_c)^2 \over (f^{(2)}_c)^3} -{1\over 8} {f^{(4)}_c \over (f^{(2)}_c)^2} \right)\\
&+\CO(N^{-3}). 
\ea
\ee
We can now read the functions $\CF_k(t)$ from this expression. Note that in order to find the correct result it is 
crucial to expand as well the function of $N$ appearing in the 
first line, 
\be
{1\over N} \log\left[ { {(N/2)^{N/2} \over \Gamma(N/2)}} {\sqrt{2 \pi \over N}} \right]={1\over 2} - {1\over 2N} \log(2) -{1\over 6 N^2} + \cdots
\ee
We conclude that 
\be
\label{1Nexpansion}
\ba
\CF_0(t)&= {1\over 2} -f_c,\\
\CF_1 (t)&= -{1\over 2 } \log(2 f''_c), \\
\CF_2(t)&=-{1\over 6}+ {5 \over 24} {(f^{(3)}_c)^2 \over (f^{(2)}_c)^3} -{1\over 8} {f^{(4)}_c \over (f^{(2)}_c)^2}, 
\ea
\ee
and so on. One can check that (\ref{1Nexpansion}) reproduces the small $t$ expansions in (\ref{small-lam}). 

We can compute easily the leading large $N$ instanton of this theory, following 
\cite{hikami-brezin}, by considering the two additional critical points of $f(\xi)$ at 
\be
\xi_B^\pm  =\log\left[ {(1+ 4 t)^{1/2}+1 \over 2 t} \right] \pm \ri \pi. 
\ee
Like before, we will denote 
\be
f^{(n)}_B=f^{(n)}(\xi^\pm_B), \qquad n \ge 1, 
\ee
since it does not depend on the choice of sign. We will also denote
\be
f^\pm_B= f(\xi^\pm_B). 
\ee
The trans-series associated to the saddle point $\xi_B^\pm$ is given by 
\be
{ {(N/2)^{N/2} \over \Gamma(N/2)}}
 {\sqrt{2 \pi \over N  f^{(2)}_B}}\re^{-N f^\pm_B} \left\{ 1+{1\over N} \left( {5 \over 24} {(f^{(3)}_B)^2 \over (f^{(2)}_B)^3} -{1\over 8} {f^{(4)}_B \over (f^{(2)}_B)^2} \right)+ \CO(N^{-2}) \right\}.
\ee
 It was argued in \cite{hikami-brezin} that this large $N$ instanton 
 controls the large order behavior of the $1/N$ expansion of the free energy (\ref{F1N}). 
 As pointed out also in \cite{hikami-brezin}, one can use this large $N$ instanton to obtain the 
 resurgent structure of the perturbative series (\ref{pfe-qvm}) at fixed $N$. In particular, 
 one can deduce the resurgent structure of the series $\mathfrak{R}(g)$ (\ref{quartic-replica}) obtained in the 
 replica limit. To see this, 
 one sets $t=gN$ and expands around $g=0$, to extract the corresponding instanton amplitude. In this paper we 
 will follow the conventions of e.g. \cite{gm-multi,amp} and define these amplitudes after dividing by $\ri$ the result of the 
 saddle-point evaluation. One finds in this way, 
\be
\label{qvm-stand-inst}
{\mathfrak{Z}^{(1)}_\pm (g;N) \over \mathfrak{Z}(g;N)}= {2^{1-N/2} \over \Gamma(N/2)}\re^{\mp \pi \ri (N- 1) /2} {\sqrt{\pi}} g^{-{N-1 \over 2} }  \re^{1 \over 4g} \exp\left[ \left({N^2 \over 2} -N +2 \right)g+ \CO(g^2) \right]. 
\ee
This gives the conventional instanton amplitude at fixed $N$. We can deduce from this amplitude 
the large order behavior of the series $\mathfrak{F}(g;N)$ in (\ref{pfe-qvm}). If we write the r.h.s. of (\ref{qvm-stand-inst}) in the standard way 
\be
g^{-b} \re^{-A/g} (c_0+ c_1 g+ \cdots)
\ee
we read off
\be
b={N-1 \over 2}, \qquad A=-{1\over 4}, \qquad c_0= {2^{1-N/2} \over \Gamma(N/2)}\re^{\mp \pi \ri (N-1)/2} {\sqrt{\pi}}, \qquad c_1= c_0  \left({N^2 \over 2} -N +2 \right).
\ee
Standard resurgent analysis leads to the following asymptotic formula for the coefficients $a_k^{(c)}(N)$ of the free energy (\ref{pfe-qvm})  
\be
\label{lo-qvm}
a_k^{(c)}(N) \sim  {2^{-N/2} \over {\sqrt{\pi}} \, \Gamma(N/2)} (-1)^k \left( {1\over 4} \right)^{-k-{N-1 \over2} } \Gamma\left(k+ {N-1 \over 2} \right)
 \left[1-{  {N^2 \over 2} -N +2 \over 4 k} + \cdots\right].
\ee
Note that the phase of $A^{-k -{N-1 \over 2}}$ in the large order formula combines with the phase of $c_0$ to give a real result. 
One can check e.g. numerically that (\ref{lo-qvm}) gives the right asymptotics of the 
series for arbitrary real values of $N$ (not necessarily integers). 
The inverse Gamma function guarantees that the asymptotics vanishes when $N=-2 s$, 
$s\in \IZ_{\ge 0}$, since for these values the coefficients $a_k(N)$ in (\ref{phiN}) 
vanish after $k=[s/2]$ and the series $\mathfrak{F}(g;N)$ is convergent. In particular, one can 
consider the replica limit of the trans-series, or equivalently, of the asymptotic formula (\ref{lo-qvm})
\be
\lim_{N \to 0}  { a_k^{(c)}(N) \over N}= {1\over 2 {\sqrt{\pi}}} (-1)^k \left( {1\over 4} \right)^{-k+{1 \over2} } \Gamma\left(k-{1 \over 2} \right)
 \left[1-{ 1 \over 2 k} + \cdots\right], 
 \ee
 which gives the correct asymptotic behavior for the series $\mathfrak{R}(g)$ in (\ref{quartic-replica}). 

The lesson to be extracted from this example is that, in theories which admit a large $N$ limit, the resurgent properties of 
the conventional perturbative series at fixed $N$ can be obtained by considering the small 't Hooft coupling limit of large $N$ instantons. 
In fact, even the replica limit of ``small" $N$ is controlled by the ``large" $N$ asymptotics!

In the case of the quartic vector model, one can also 
obtain the standard instanton amplitude (\ref{qvm-stand-inst}) by 
doing a conventional saddle-point calculation directly in the defining integral (\ref{zng}). However, in 
order to obtain the correct result, one has 
to integrate carefully over collective coordinates due to the broken $O(N)$ symmetry. The large $N$ calculation 
incorporates automatically this integration and is more convenient. We will now consider an example 
where the only way we know of obtaining the standard instanton amplitude is to appeal to large $N$ instantons.

\subsection{Chern--Simons theory on the three-sphere}

As is well-known, 
$U(N)$ Chern--Simons theory is a topological gauge theory for which many observables can be computed in closed form, see \cite{mmcsts} for a review with relevant information for the contents of this section. We will consider the free energy on the three-sphere $\IS^3$, which is given by \cite{witten-cs} 
\be
\label{np-cs}
\mF(g_s;N) =\sum_{j=1}^N (N-j) \log\left[ {2 \over j g_s} \sin \left ( {j g_s \over 2} \right) \right]. 
\ee
Here, $g_s$ is the coupling constant. It is related to the Chern--Simons level $k$ by 
\be
g_s={2 \pi \over k+N}. 
\ee
The expression (\ref{np-cs}) is the non-perturbative, exact result for any positive integer $N$. From this exact result 
one can easily deduce the perturbative series. It is convenient to introduce the coupling $\lambda$ as 
\be
 g_s=\ri \lambda. 
 \ee
 Let us define
\be
\label{fgj}
\ba
f_{0,j}&= {B_{2j} \over 2 j (2j+2)!}, \\
f_{1,j}&= -{1\over 12}{  B_{2j} \over 2j (2j)!}, \\
f_{g,j}&= -{1\over (2j+2-2g)!} {B_{2j} B_{2g} \over 4j g (2g-2)!},  \qquad g \ge 2, 
\ea
\ee
for $j\ge 1$. Here, $B_{2k}$ are Bernoulli numbers. Then, one has
\be
\label{ptcs}
\mathfrak{F}(g_s; N)=\sum_{j=1}^\infty f_j (N) \lambda^{2j}, \qquad f_j(N)= \sum_{g=0}^j f_{g, j} N^{2j+2-2g}. 
\ee
 In contrast to the original expression (\ref{np-cs}), this perturbative result is 
well-defined, as a formal power series, for any complex value of $N$. In addition, it 
has a peculiar behavior. When $N$ is a non-zero integer, the series has a finite radius of 
convergence. This is as it should be, since when $N$ is a positive integer, this series is the expansion of (\ref{np-cs}), which is an analytic function of $g_s$ at the origin. In addition, the coefficients $f_j(N)$ are even functions of $N$, therefore we have the
same radius of convergence for negative, integer $N$. However, when $N$ is not an integer, the series turns 
out to be factorially divergent, 
like generic perturbative series in QFT. One can also obtain a replica limit by extracting an overall $N^2$ factor, 
\be
\label{cs-replica}
\mathfrak{R}(g_s)=\lim_{N \to 0} {\mathfrak{F}(g_s; N) \over N^2}=\sum_{n \ge 1} r_n \lambda^{2n}, \qquad r_n= -{B_{2n}^2 \over 8 n^2 (2n-2)!}, 
\ee
which is also given by a factorially divergent series. 

We can now ask what are the resurgent properties of the perturbative series (\ref{ptcs}). Equivalently, 
what are the relevant instanton trans-series, and what is the asymptotic behavior of the coefficients in (\ref{ptcs})? Note that this question has only a non-trivial answer when $N$ is not a non-zero integer, and we expect the trans-series to vanish for integer $N$. 
It is not obvious how to answer this question directly in Chern--Simons theory, since the relevant instantons should only appear when $N$ is not an integer, i.e. precisely when the theory is not defined by its standard gauge theory construction. 
Following the insights of the toy model in the previous section, we consider the theory at large $N$ and its large $N$ instantons. 

The 't Hooft expansion of $U(N)$ Chern--Simons theory on $\IS^3$ was found in \cite{gv-cs}. We introduce the 't Hooft coupling  
\be
t= N \lambda
\ee
so that we have the following $1/N$ expansion, 
\be
\label{cs-largeN}
\CF(t;g_s)= \sum_{g \ge 0} \CF_g(t) g_s^{2g-2}. 
\ee
 The functions $\CF_g(t)$ are explicitly given by
\be
\label{fg-sphere}
\ba
\CF_0(t)&={\rm Li}_3 (\re^{-t}) -{t^2 \over 4} \left(3-2 \log(t) \right)- \zeta(3) + {\pi^2 t \over 6} -{t^3 \over 12}, \\
\CF_1(t)&= {1\over 12} {\rm Li}_1 (\re^{-t}) +{1\over 12} \log(t)-{t\over 24}, \\
\CF_g(t)&= c_g +  { (-1)^{g-1}B_{2g} \over 2g (2g-2)!}  \left( {\rm Li}_{3-2g} (\re^{-t})-(2g-3)! t^{2-2g} \right), \qquad g\ge 2, 
\ea
\ee
where 
\be
c_g= 2 {(-1)^{g-1} B_{2g} B_{2g-2} \over 4 g (2g-2) (2g-2)!}. 
\ee
The free energies $\CF_g(t)$ appearing in (\ref{fg-sphere}) are slightly different from the ones usually 
considered in the topological string literature: we have removed from them a singular part due to the conifold behavior, as well 
as the so-called constant map contribution. In this way, after setting $t=N \lambda$, one recovers the 
perturbative series (\ref{ptcs}). The subtraction of these two ingredients was already part of the algorithm in \cite{gm-peacock} to obtain 
perturbative series from topological string free energies, and we will implement it in all the examples considered in this paper, as it will 
become clear in the next sections.

\begin{figure}
\centering
\includegraphics[width=0.35\textwidth]{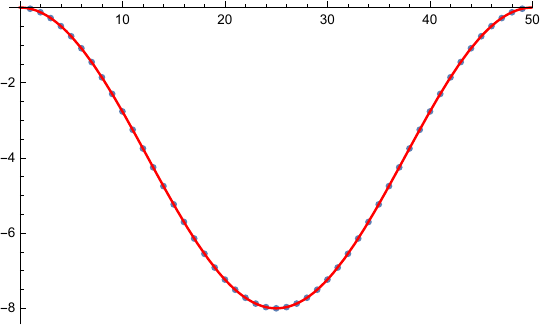}
\hspace{0.065\textwidth}
\includegraphics[width=0.35\textwidth]{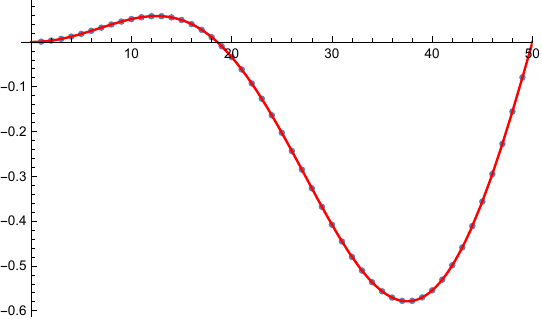}
\caption{The dots represent the numerical results for $\mu_{0,1}$ (left and right, respectively) for the values of $N$ of the form $i/50$, $i=1, \cdots, 49$, 
while the lines are the theoretical predictions (\ref{preds}).}
\label{fig-csloplots}
\end{figure}

The multi-instanton amplitudes of the large $N$ theory were found in \cite{ps09}. They have the Pasquetti--Schiappa form 
\be
\label{tia}
\CF_\CA^{(\ell)}= a \left[ {1 \over \ell}  \left( {\CA \over g_s} \right) + {1 \over \ell^2} \right] \re^{-\ell \CA/g_s}, \qquad \ell \in \IZ_{>0}, 
\ee
where 
\be
\label{aS}
a= { \Omega \over 2 \pi}, 
\ee
and in this case $\Omega=1$ (in the more general case, $\Omega$ 
will be a non-trivial Stokes constant, 
conjecturally related to a counting of BPS states). The instanton action $\CA$ appearing in 
(\ref{tia}) can take the values
\be
\label{res-con-action}
\CA_{\pm, k}=\pm  2 \pi (t + 2 \pi \ri k), \qquad k \in \IZ.  
\ee
The leading amplitude corresponds to $\ell=1$ in (\ref{tia}) and $k = \pm 1$ in (\ref{res-con-action}) 
(the value $k=0$ leads to a vanishing action in the perturbative limit, and is not relevant for us). 
When we express this amplitude in terms of $\lambda$, 
and we add the contributions of $\CA_{+,1}$ and $\CA_{-, -1}$, we find 
\be
{1 \over  \pi \lambda} \re^{- 4 \pi^2/\lambda} \left\{ 4 \pi^2 \cos \left( 2 \pi N \right)+ \lambda \left( \cos \left( 2 \pi N \right)+ 2 \pi N \sin \left( 2 \pi N \right) \right) \right\}.
\ee
In order to obtain the relevant trans-series for the perturbative series 
(\ref{ptcs}), which vanishes quadratically as $N \to 0$, we have to subtract the value of the amplitude 
at $N=0$ (the value at $N=0$ is related to so-called constant map contribution to the free energy $\CF_g(t)$, which is independent of 
$t$ and that we have subtracted). We finally obtain, 
\be
\label{trans-sub}
\mathfrak{F}^{(1)}(\lambda;N)= {1\over  \pi \lambda} \re^{- 4 \pi^2/\lambda} \left\{ 4 \pi^2 \left(\cos \left( 2 \pi N \right)-1 \right)+ \lambda \left( \cos \left( 2 \pi N \right)-1+ 2 \pi N \sin \left( 2 \pi N \right) \right) \right\}. 
\ee
This leads to the asymptotic behavior 
\be
\label{fcs-as}
f_n (N) \sim (4 \pi^2)^{-2n -1} \Gamma (2n+1) \left\{ \mu_0(N) +  { 4 \pi^2  \over 2 n} \mu_1(N) \right\}, \qquad n \gg 1, 
\ee
where 
\be
\label{preds}
\mu_0(N)= 4 \left(\cos \left( 2 \pi N \right)-1 \right), \qquad \mu_1(N)={1 \over \pi^2} \left( \cos \left( 2 \pi N \right)-1+ 2 \pi N \sin \left( 2 \pi N \right) \right). 
\ee
Note that, if $N$ is a non-zero integer, $\mu_{0,1}(N)$ vanish, and we recover the fact that the series $f_j(N)$ has a finite 
radius of convergence. 
The asymptotic behavior (\ref{fcs-as}) can be checked numerically in detail. In \figref{fig-csloplots} we show the numerical results for $\mu_{0,1}(N)$ for values of 
$N=i/50$, $i=1, \dots, 49$, and we compare them with the analytic prediction. 

In particular, one can  extract from (\ref{fcs-as}) the asymptotics of the perturbative series in the replica limit, by simply extracting the leading coefficient of (\ref{trans-sub}) as $N \to 0$, and 
one finds 
\be
r_n \sim (4 \pi^2)^{-2n -1} \Gamma (2n+1) \left\{ -8 \pi^2 +  {4 \pi^2  \over n} \right\}, \qquad n \gg 1, 
\ee
which can be verified analytically. 

The example of Chern--Simons theory on $\IS^3$, although very simple, illustrates 
various aspects of the general situation. First of all, it gives yet another instance of the general idea that 
large $N$ instantons are useful to understand the asymptotics of the perturbative series for 
arbitrary values of $N$. It also shows that the values of $N$ for which the theory was originally defined (positive integers) 
are very special from the point of view of the resurgent structure, and lead to a simplification of the trans-series (in this case, 
to its vanishing). We will encounter these properties in the more general setting of matrix models and topological strings, which 
we address in the following sections.

\sectiono{General formalism for matrix models and topological strings}
\label{sec-gt}

We will now study how to obtain conventional instanton amplitudes at fixed $N$ from large $N$ instanton amplitudes, 
in the case of matrix models and topological strings. For simplicity, we will consider models with one single modulus, although the generalization to a multi-modulus case should be straightforward. In the 
case of matrix models, this means that we are considering models with one or two-cuts in their eigenvalue distribution. 
We will consider first the one-cut matrix model, since it is simpler, and then we will consider the generic one-modulus case. 
For background on matrix models and their $1/N$ expansion, see e.g. \cite{mmbook, mmhouches}. 

Let us first set up some general features common to matrix models and topological strings with 
one modulus. In these theories, the $1/N$ expansion of the free energy reads
\be
\CF (t; g_s)= \sum_{ g \ge 0} g_s^{2g-2} \CF_g(t), 
\ee
where $t=g_s N$ is the 't Hooft coupling. We will assume 
that the free energies $\CF_g(t)$ have a conifold 
behavior when $t \rightarrow 0$, i.e. they have the form,   
\be
\label{fg-cr}
\CF_g(t)= \CF_g^{\rm con} (t) + \CF^{\rm reg}_g(t).  
\ee
Here, $\CF_g^{\rm con}(t)$ are the conifold free energies. For $g=0,1$ one has
\be
\label{con-free}
\CF^{\rm con}_0(t)= {t^2 \over 2} \left( \log(t)-{3 \over2} \right), \qquad \CF^{\rm con}_1(t)=
 -{1\over 12} \log(t),\ee
while for general $g\ge 2$ we obtain 
\be
\label{fg-con}
 \CF^{\rm con}_g(t)= {B_{2g} \over 2g (2g-2)} t^{2-2g}, \quad g\ge 2. 
 \ee
The free energies $\CF^{\rm reg}_g(t)$ are analytic at $t=0$. In the 
case of matrix models, the structure (\ref{fg-cr}) is a direct consequence of the perturbative expansion. 
In the case of topological strings, the structure (\ref{fg-cr}) is typical of the free energies in the so-called 
conifold frame \cite{gv-conifold}. Therefore, when considering the topological string, we will assume that we are in that frame, unless we 
explicitly state otherwise. An important aspect of the structure (\ref{fg-cr}) for $g \ge 2$ is that in the Laurent expansion around $t=0$ there are no terms in $t^{-n}$ with $1\le n\le 2g-1$. This is sometimes called the {\it gap property} or the {\it gap condition}. 

One can also expand the free energy as a series in the coupling constant $g_s$, 
at fixed $N$. This expansion has the form
\be
\label{oc-pert}
 \mfF(g_s;N)= \sum_{n \ge -1} c_n (N) g_s^{n}.   
\ee
It can be obtained by setting $t=N g_s$ in 
\be
\CF^{\rm reg}(t;g_s)= \sum_{ g \ge 0} g_s^{2g-2} \CF^{\rm reg}_g(t). 
\ee
The coefficients $c_n(N)$ in (\ref{oc-pert}) can be written as
\be
c_n(N)= \sum_{2g+h-2=n} f_{g,h} N^h,
\ee
where $f_{g,h}$ are the coefficients in the expansion of the genus $g$ free energies around $t=0$:
\be
\label{thooft-undone}
\CF^{\rm reg}_g(t)= \sum_{h \ge 0} f_{g,h} t^h. 
\ee
Note that $c_n(N)$ is a polynomial in $N$, and it satisfies
\be
\label{symc}
c_n(-N)= (-1)^n c_n(N). 
\ee
The replica limit is in this case
\be
\label{r-oc}
\mathfrak{R}(g_s)= \lim_{N \to 0} { \mathfrak{Z}(g_s;N) -1 \over N}= \sum_{g \ge 0} f_{g,1} g_s^{2g-1}, 
\ee
and we have 
\be
f_{g,1}= \lim_{N\to 0} {c_{2g-1}(N) \over N}. 
\ee
The replica limit captures all-genus information, but it keeps only the first term in the expansion of the free energies. 

\subsection{The one-cut matrix model}

We want to obtain information on the resurgent structure of the series (\ref{oc-pert}) and its replica limit (\ref{r-oc}), 
in the case of the one-cut matrix model, by 
using its large $N$ instantons. These were studied in \cite{msw} in the one-instanton case.  
General multi-instanton amplitudes for models with one-modulus were obtained in \cite{multi-multi}. The idea of \cite{multi-multi} 
was to consider the two-cut generalization of the one-cut case, and take the limit in which the spectral curve of the model 
is pinched at one cycle. In the two-cut case there are two different 't Hooft parameters, $t_i= g_s N_i$, $i=1,2$, where $N_1$, $N_2$ are the number of eigenvalues in the matrix integral which are expanded around the two critical points 
(see e.g. \cite{marmir} for a summary of the two-cut case). The two-cut, genus $g$ free energies have the structure
\be
\label{decom}
\CF_g(t_1, t_2)= \widehat{\CF}_g(t_1, t_2) + \CF_g^{\rm con}(t_2)= \CG_g(t_1, t_2)+ \CF_g^{\rm con}(t_1)+\CF_g^{\rm con}(t_2), 
\ee
where the second equality defines the functions $\CG_g(t_1, t_2)$ introduced in \cite{marmir}. 
The function $\widehat{\CF}_g(t_1, t_2)$ is regular as 
$t_2 \rightarrow0$, while $\CG_g(t_1, t_2)$ is regular when both $t_1, t_2 \rightarrow 0$. The formula of 
\cite{multi-multi} for the large $N$ $\ell$-instanton amplitude is 
 \be
 \label{msw-formula}
 Z^{(\ell)}_{\rm r}= Z^{(\ell)}_{\rm G} \exp \left\{ \sum_{g \ge 0}  \left( \widehat{\CF}_g(t- \ell g_s,\ell g_s) - \CF_g(t)\right) g_s^{2g-2} \right\}, 
 \ee
where $\CF_g(t)= \widehat{\CF}_g(t, 0)$ is the genus $g$ free energy of the one-cut matrix model. 
In this formula we are considering ``reduced" partition functions, 
 \be
 Z^{(\ell)}_{\rm r}= {Z^{(\ell)} \over Z^{(0)}}. 
 \ee
They are normalized by dividing by the zero-instanton partition function, which gives the conventional 
 $1/N$ expansion. The prefactor in (\ref{msw-formula}) is given by
 \be
 \label{gaussian}
 Z^{(\ell)}_{\rm G} = {g_s^{\ell^2/2} \over (2 \pi)^{\ell/2}} G_2(\ell+1). 
 \ee
 It is the exact partition function of the Gaussian matrix model, and it regulates the conifold 
 free energies $\CF^{\rm con}_g(t_2)$ in (\ref{con-free}). The physical meaning of (\ref{msw-formula}) is simply that 
 multi-instantons in the one-cut matrix model can be obtained by eigenvalue tunneling \cite{david, shenker}, 
 in which a small amount $\ell \ll N$ of eigenvalues in the cut ``tunnel" to the other critical point. 
 
According to the logic of the examples in the previous section, the 
standard instanton amplitudes should be obtained by simply setting $t=N g_s$ 
in the multi-instanton formula (\ref{msw-formula}), and expanding around $g_s=0$. If we do this, we find a new 
 singular part, this time coming from $\CF^{\rm con}_g(t -\ell g_s)- \CF^{\rm con}_g(t)$, which we have to regulate. This is in contrast to the examples we discussed in section \ref{sec-warmup}, since the perturbative limit is not completely straightforward and we have to find the appropriate regularization. Inspired by the procedure in \cite{multi-multi}, we propose the regularization
 \be
 \label{reg-fgl}
 \ba
 \CF^{\rm con}(t- \ell g_s)- \CF^{\rm con}(t)  \longrightarrow& {\ell^2 \over 2} \log(g_s) +{\ell \over 2} \log(2 \pi)- \ell N \log(g_s)
 \\ 
  +& \log\, G_2 (N+1-\ell) - \log \, G_2(N+1). 
  \ea
 \ee
The rationale for this procedure is that, when $t=Ng_s$, the l.h.s. of (\ref{reg-fgl}) is a series in $1/N$ which agrees with the asymptotic expansion of the r.h.s. Setting $t=N g_s$ in the contribution of the regular part, encoded in the functions $\CG_g(t_1, t_2)$, is straightforward. 
By expanding in the 
 't Hooft parameters we find, 
 \be
 \CG_g(t_1, t_2)= \sum_{h_1, h_2} \CG_{g, h_1, h_2} t_1^{h_1} t_2^{h_2}, \qquad  \CG_g(t)= \sum_{h} \CG_{g, h} t^{h}, 
 \ee
 where $\CG_g(t)= \CG_g(t,0)$. Let us define 
 \be
 f_m^{(\ell)}(N)= -\sum_{g, h_1, h_2 \atop 2g-2+h_1+h_2=m} \CG_{g, h_1, h_2} (N-\ell)^{h_1} \ell^{h_2} + \sum_{g, h \atop 2g-2+h=m} \CG_{g, h}N^h. 
 \ee
 We note that the genus zero contribution has the structure
 \be
 \label{g0exp}
\CG_0 (t_1, t_2) = {\CA \over 2} (t_1- t_2)+ \cdots, 
\ee
where the dots stand for higher powers of the 't Hooft parameters. Then, after 
exponentiation and multiplication by $ Z^{(\ell)}_{\rm G}$, we obtain the conventional $\ell$-instanton amplitude 
 \be
 \label{one-cut-ell}
 \mathfrak{Z}^{(\ell)}(g_s;N)= g_s^{\ell(\ell- N)}  { G_2(\ell+1) \over \prod_{k=0}^{\ell-1} \Gamma(N-k) }  \re^{-\ell { \CA \over g_s}} \exp\left(- \sum_{m \ge  0} f_m ^{(\ell)} (N) g_s^m \right), 
 \ee
 where we have used the defining property of Barnes function, 
 \be
 G_2(N+1)= \Gamma(N) G_2(N). 
 \ee
  We see that $\CA$ in (\ref{g0exp}) has the interpretation of an instanton action for the perturbative series. 
  Note that, due to the presence of the inverse Gamma functions, for a fixed integer $N$ 
  only the multi-instanton amplitudes with $1\le \ell \le N$ are non vanishing. A particularly important case of the general 
  formula above occurs for the one-instanton case $\ell=1$, 
 \be
 \label{lone-ex}
\mathfrak{Z}^{(1)}(g_s;N)=  { 1 \over \Gamma(N)} g_s^{1- N} \re^{-{\CA \over g_s}}\exp\left(- \sum_{m \ge 0} f_m ^{(1)} (N) g_s^m \right). 
 \ee
 When $N$ is a negative integer or zero, this amplitude vanishes due to the inverse Gamma function. This is similar to what 
 happened to the instanton (\ref{qvm-stand-inst}) in the $O(N)$ model. 
 It is interesting to consider the replica limit in which we take $N \to 0$, after dividing by $N$. In this limit, only the one-instanton 
 amplitude survives, and we find 
 \be
\mathfrak{Z}_R ^{(1)} (g_s)=\lim_{N \to 0} {\mathfrak{Z}^{(1)}(g_s;N) \over N}= g_s \, \re^{-{\CA \over g_s}}\exp\left(- \sum_{m \ge 0} f_m ^{(1)} (0) g_s^m \right).  
 \ee

In order to obtain a complete description of the asymptotics of the 
conventional perturbative series we have to include ``negative" instantons, see e.g. \cite{gikm,asv11,mss}. Their presence 
is necessary due to the symmetry (\ref{symc}), which implies that the 
asymptotic formula for the coefficients $c_k(N)$ must be of the form 
\be
\label{ckN}
c_k(N) \sim f_k(N)+ (-1)^k f_k (-N), 
\ee
for appropriate functions $f_k(N)$. One of the two terms in the r.h.s. of (\ref{ckN}) comes from positive instantons, while 
the other comes from negative instantons. A formula for negative instanton amplitudes can be obtained as follows. The result of \cite{multi-multi} can be extended to negative instantons as follows \cite{mss}
 \be
 Z^{(0|\ell)}= Z^{(\ell)}_{\rm G} \exp \left\{ \sum_{g \ge 0}  \left( \widehat{\CF}_g(t+\ell g_s,-\ell g_s) - \CF_g(t)\right) g_s^{2g-2} \right\}. 
 \ee
 Like before, when we set $t= N g_s$ and expand around $g_s=0$, there are singular terms which have to be regulated. 
We propose the following regularization:
 \be
 \label{reg-plus}
 \ba
 \CF^{\rm con}(t+ \ell g_s)- \CF^{\rm con}(t) \longrightarrow &{\ell^2 \over 2} \log(g_s) +\ell N \log(g_s)+ {\ell \over 2}  \log(2 \pi) -\ri \pi \ell N  \\ 
 - & {\ri \pi \over 2} \ell^2+ \log\, G_2 (-N+1-\ell) - \log \, G_2(-N+1).
 \ea 
 \ee
 This has the property that the asymptotic expansion of the r.h.s. reproduces the l.h.s., and in addition it leads to a result compatible 
 with (\ref{ckN}). By using this regularization one obtains the conventional instanton amplitude,
 \be
 \mathfrak{Z}^{(0|\ell)}(g_s;N)= \re^{-\pi \ri \ell N} \ri^{-\ell^2} g_s^{\ell(\ell+ N)} {  G_2(\ell+1) \over \prod_{k=0}^{\ell-1} \Gamma(-N-k)} \re^{{ \CA  \ell \over g_s}} \exp\left(- \sum_{m \ge 0} f_m ^{(\ell)} (-N) (-g_s)^m \right). 
 \ee
In the case $\ell=1$, we have
 \be
 \label{lone-antiex}
 \mathfrak{Z}^{(0|1)}(g_s;N)= -\ri  { \re^{-\ri \pi N}  \over \Gamma(-N)} g_s^{1+ N} \re^{{\CA \over  g_s}}\exp\left(- \sum_{m \ge 0} f_m ^{(1)} (-N) (-g_s)^m \right). 
 \ee
Note that, when $N$ is a non-negative integer, this amplitude vanishes due to the inverse Gamma function.

\subsection{Multi-cut matrix models and topological strings}

We consider now large $N$ instantons in multi-cut matrix models and topological 
strings on local Calabi--Yau manifolds. They are 
described by a common formalism \cite{gm-multi,marmir} obtained from trans-series solutions to the holomorphic anomaly 
equations \cite{cesv1, cesv2}. 

Let us first present results for one-instanton amplitudes. As we mentioned above, 
we will focus on models with a single modulus. The corresponding 
flat coordinate will be denoted by $t$. In the case of matrix models, $t$ can 
be interpreted as a 't Hooft parameter. In a two-cut matrix model 
with a single modulus, there are two 't Hooft parameters $t_{1,2}$, as we mentioned before. Their sum 
$t_1+t_2=T$ can be regarded as a ``mass parameter," and we can take $t=t_1$. 
The genus $g$ free energies will be denoted by $\CF_g(t)$. In the two-cut case, we will sometimes write them as
 $\CF_g(t_1, t_2)$ or $\CF_g(t;T)$, depending on the context. 
 
It has been argued in \cite{kazakov-kostov,dmp-np,cesv1, cesv2,gkkm} that the possible instanton actions are 
integer linear combinations of periods. They take the form 
\be
\label{acr}
\CA=  c { \partial \CF_0 \over \partial t} +2 \pi \ri r t+ 4 \pi^2 \ri n,  
\ee
where $c,r, n \in \IZ$, for appropriate normalizations of $t$ and $\CF_0$. If $c$ vanishes, the multi-instanton amplitudes 
have the Pasquetti--Schiappa form (\ref{tia}), where the coefficient $a$ involves in general a Stokes 
constant $\Omega$. It has been conjectured \cite{mm-s2019,gm-peacock,astt, gm-multi, ghn,aht,gu-res,gkkm,amp} 
that these Stokes constants are given by BPS invariants, and we will use this interpretation in various contexts. 

It is sometimes useful to put all these multi-instanton amplitudes 
together in a formula for the Stokes automorphism, as it was done in \cite{im}. The Stokes automorphism can 
be defined through the alien derivatives, by \'Ecalle's formula 
(see e.g. \cite{msauzin})
\be
\mathfrak{S}_\CC = \exp \left( \sum_{\ell =1}^\infty \CC^{\ell} \dot \Delta_{\ell \CA}  \right), 
\ee
where the additional formal parameter $\CC$ has been introduced to keep 
track of the $\ell$-th instanton sector. We can then write
\be
\label{stokes-afr}
\ba
\mathfrak{S}_{\CC} (Z^{(0)})&=\exp \left( \ri \sum_{\ell =1}^\infty  \CC^{\ell} \CF^{(\ell)}_\CA \right)Z^{(0)} 
\\ &=
\exp \left\{  \ri a \left( {\rm Li}_2\left(\CC \, \re^{-\CA/g_s}\right)-  {\CA  \over g_s} \log\left(1-\CC \, \re^{-\CA/g_s}\right) \right) \right\}Z^{(0)}, 
\ea
\ee
where $a$ is given in (\ref{aS}). 

If $c$ does not vanish, we define a modified prepotential $\CF^\CA_0$ by
\be
\label{mod-prep}
\CA=c {\partial \CF^\CA_0 \over \partial t}. 
\ee
Explicitly, we have
\be
\CF^\CA_0 (t)= \CF_0 + {\pi \ri r \over c} t^2 + {4 \pi^2 \ri n \over c} t. 
\ee
Then, the one-instanton 
amplitude associated to the action $\CA$ is given by \cite{gm-multi}
\be
\label{ntinst}
\CF^{(1|0)}= a 
 \left(1+ g_s c {\partial \CF \over \partial t} ( t- c g_s; g_s) \right)  \exp \left[ \CF(t-c g_s; g_s)-\CF(t; g_s) \right]. 
\ee
Here, $\CF$ is the total free energy, in which $\CF_0$ has been replaced by the modified 
prepotential $\CF^\CA_0$. We note that (\ref{ntinst}) is given by a non-trivial 
formal power series in $g_s$, 
\be
  \label{ex-f1}
  \ba
  \CF^{(1|0)}
&=a \,\re^{-\CA/g_s} \exp\left( {c^2 \over 2}\partial_t^2 \CF_0 \right)\\
  &\qquad \times 
  \left\{ {\CA  \over g_s} + 1 -c^2 \partial_t^2 \CF_0 - \CA \left( c \partial_t \CF_1 + {c^3 \over 6}  \partial^3_t \CF_0 \right) + \CO(g_s)  \right\}. 
  \ea
  \ee
The anti-instanton amplitude is
\be
\label{ntainst}
\CF^{(0|1)}=a 
\left(1- g_s c {\partial \CF \over \partial t} ( t+c g_s; g_s) \right)  \exp \left[ \CF(t+c g_s; g_s)-\CF(t; g_s) \right]. 
\ee

The multi-instanton amplitudes are given by slightly more complicated formulae, but a simple 
generating function for them was obtained in \cite{im}. It is better expressed in terms 
of $\ell$-instanton partition functions. Its explicit form is 
\be
\label{stokesZ}
\mathfrak{S}_{\CC} (Z^{(0)})= \sum_{\ell \ge 0} \CC^\ell Z^{(\ell)}(t)= 
\sum_{\ell \ge 0} \CC^{\ell} z^{(\ell)}  (t- \ell \hbar) Z(t- \ell \hbar) , 
\ee
where 
\be
\hbar= c g_s. 
\ee
When no superscript is indicated, we assume that the free energy or the partition functions are the perturbative ones. 
The functions $z^{(\ell)}(t)$ appearing in (\ref{stokesZ}) can be obtained from 
\be
\label{simple-identity}
 \re^{\ri a {\rm Li}_2(\CC)}  {Z\left(t - \hbar  \ri a \log(1-\CC) \right) \over Z(t)}= \sum_{\ell \ge 0}  \CC^{\ell} z^{(\ell)}  (t).  
 \ee
One finds, for example, 
\be
\label{multi-ex}
\ba
\mathfrak{S}_{\CC} (Z^{(0)})
&=Z^{(0)}+ \ri a  \CC \left( 1+ \CF'_{[-1]}   \right)  \re^{ \CF_{[-1]}}- { a^2 \CC^2  \over 2}  \left(1+2  \CF'_{[-2]} +  (\CF'_{[-2]} )^2+  \CF'' _{[-2]} \right)  \re^{ \CF_{[-2]}}\\
&+ 
\ri a \CC^2 \left( {1\over 4} + {\CF'_{[-2]} \over 2} \right)  \re^{ \CF_{[-2]}}+ \cdots
\ea
\ee
where we used the notation  
\be
\CF_{[k]}(t)= \CF(t+k \hbar)
\ee
and we set $\hbar=1$ (one can restore the factors of $\hbar$ by simply taking into account that each 
derivative has one of these factors). The multi-instanton free energies are then defined, as in (\ref{stokes-afr}), by 
\be 
\sum_{\ell =1}^\infty \CC^\ell \CF^{(\ell|0)}= -\ri \log \left( {\mathfrak{S}_{\cal{C}} (Z^{(0)}) \over Z^{(0)}} \right). 
\ee
One can extend this formula to define more general amplitudes $\CF^{(\ell_1| \ell_2)}$ \cite{im}, but 
we will not need them in this work. 

We will now study these multi-instanton amplitudes when we set $t= Ng_s$ and we expand in $g_s$. We will first consider the simplest case, namely when $c=0$ in (\ref{acr}) and the 
instanton amplitudes have the form (\ref{tia}). The instanton actions are of the form 
\be
\label{arn}
\CA_{r, n} = 2 \pi \ri r t + 4 \pi^2 \ri  n, \qquad r,n \in \IZ, 
\ee
and the corresponding Stokes constants are denoted by $\Omega_{r,n}$. After setting $t= N g_s$, we find 
\be
\label{psN}
\ba
&\sum_{r,n} \mathfrak{F}^{(r,n)}(g_s;N)\\
&=
{1 \over 2 \pi }  \sum_{r, n} \sum_{\ell \ge 1}\Omega_{r,n} \re^{-{4 \pi^2 \ri \ell n \over g_s}} \left[ \left( \re^{-2 \pi \ri r N \ell} -1 \right) 
\left(  {4 \pi^2 \ri n \over g_s \ell} + {1\over \ell^2} \right) +{2 \pi  \ri r N \over \ell} \re^{-2 \pi \ri r N \ell} 
 \right],
\ea
\ee
where we summed over all possible instanton sectors labelled by $r, n$, 
and we subtracted the value of the amplitude for $N=0$, 
similarly to what was done in the case of Chern--Simons theory on $\IS^3$, in (\ref{trans-sub}). 
The expression (\ref{psN}) simplifies 
further when $N$ is an integer, and we find  
\be
\label{iN}
\sum_{r,n} \mathfrak{F}^{(r,n)}(g_s;N)=\ri N \sum_{k } R_k  \re^{-{4 \pi^2 \ri k \over g_s}}, 
\ee
where 
\be
\label{Romega}
R_k= \sum_{n \ell= k} {\Upsilon_n \over \ell}, \qquad \Upsilon_n= \sum_r r \Omega_{r,n}, 
\ee
and we have assumed that $n\ge 1$ is a positive integer (a similar expression can be of course obtained if $n$ is negative). 
As we will see later on, this simple derivation of the conventional instanton amplitudes corresponding to Pasquetti--Schiappa multi-instantons explains some of the results found in \cite{gm-peacock}. We also note that the replica limit of the instanton amplitude (\ref{psN}) is given by 
\be
\label{replica-form}
\lim_{N \to 0}{1\over N}  \sum_{r,n} \mathfrak{F}^{(r,n)}(g_s;N)= {4 \pi^2   \over g_s} \sum_{k} k R_k \,  \re^{-{4 \pi^2 \ri k \over g_s}}. 
\ee

Let us now consider the more complicated case in which $c\not=0$. We will first consider the one-instanton amplitude (\ref{ntinst}), after setting $t=Ng_s$ and expanding around $g_s=0$. 
Due to the conifold behavior, the amplitude is singular and we have to regulate it. For the exponent, this is as in 
(\ref{reg-fgl}):
 \be
 \ba
 \CF^{\rm con}(t- c g_s;g_s)- \CF^{\rm con}(t;g_s)  \longrightarrow& {c^2 \over 2} \log(g_s) +{c \over 2}  \log(2 \pi)-  cN \log(g_s)
 \\ 
  +&  \log\, G_2 (N+1-c) -  \log \, G_2(N+1). 
  \ea
 \ee
We also have to regulate the expression $g_s \partial_t \CF^{\rm con}(t-c g_s;g_s)$, which appears in the prefactor. We propose the regularization
\be
g_s \partial_t \CF^{\rm con}(t-c g_s;g_s) \rightarrow \partial_N \log G_2(N+1-c) + \log(g_s)(N-c)-{1\over 2} \log(2 \pi). 
\ee
By expressing the derivative of the Barnes function in terms of the digamma 
function $\psi(z)=\Gamma'(z)/\Gamma(z)$, we finally obtain the following 
formula for the one-instanton amplitude at fixed $N$:
  \be
  \label{f10s}
\ba
  \mfF^{(1|0)}(g_s;N)& =a (2 \pi)^{c/2} {g_s^{c(c-2N)/2} \over \prod_{k=0}^{c-1}\Gamma(N-k)}
\exp\left[\CF^{\rm reg}((N-c)g_s;g_s)-\CF^{\rm reg}(Ng_s;g_s)\right]\\
&\quad \times \re^{-2\pi\ri rN+\pi\ri rc-{4\pi^2\ri n\over g_s}} \Biggl[1-{c\over 2}+c(N-c)\left(\log(g_s)+\psi(N+1-c)-1\right)\\
&\quad \hphantom{\times \re^{-2\pi\ri rN+\pi\ri rc-{4\pi^2\ri n\over g_s}}\Biggl[}{}+c g_s\partial_t\CF^{\rm reg}((N-c)g_s;g_s)
+2\pi\ri r(N-c)+{4\pi^2\ri n\over g_s}\Biggr].
\ea
\ee
Here we have assumed that $c \in \IZ_{>0}$. The regular free energy $\CF^{\rm reg}$ is constructed from the conventional prepotential $\CF_0$, as in (\ref{fg-cr}); the contribution of $\CF_0^\CA-\CF_0$ is displayed explicitly in (\ref{f10s}). We emphasize that in obtaining this formula we have to use a regularization prescription. We will later check in examples that this formula describes e.g. the large order behavior of the perturbative expansion in $g_s$ but at fixed $N$. An important feature of 
the above one-instanton formula is that, in contrast to the original large $N$ formulae, it involves a logarithm of the coupling 
constant, $\log(g_s)$. This has implications for the resulting large order behavior, as in \cite{gikm}.  We also note 
that, in (\ref{f10s}), the terms involving $\CF^{\rm reg}(t;g_s)$ have to be expanded as a series in $g_s$ after setting $t=N g_s$. 
For example, $\CF^{\rm reg}((N-c)g_s; g_s)$ gives the perturbative series that we have denoted by $\mathfrak{F}(g_s;N-c)$.

Let us now work out a slightly more convenient formula for the one-instanton amplitude. We will fix the value of $c$ to its minimum value. This value depends on the geometry or model under consideration. For example, for topological string on local $\IP^2$ or local $\IF_0$, we have $c=3$ or $c=2$, respectively, while for the two-cut cubic matrix model we have $c=1$.

Since we are considering the limit in which $t \rightarrow 0$, the instantons with action (\ref{acr}) and different values of $r$ have to be considered together. Each of them comes with a Stokes constant which will be denoted by $\Omega_{r, n}$ (the dependence on $c$, which has been set to its minimal value, is not indicated). Then, by collecting the dependence on $r$, we have a prefactor 
\be
\label{omegaN}
\Omega_n (N)= \sum_r \Omega_{r,n} \re^{-2 \pi \ri r N + \pi \ri r c}. 
\ee
The phase in (\ref{omegaN}) is the $r$-dependent factor displayed explicitly in (\ref{f10s}). We assume in 
addition that the sum over $r$ is finite, which we have checked in examples, 
as we will see later on in the paper. When $N$ is an integer, $\Omega_n(N)$ is independent of $N$ and given by  
\be
\label{omega-int}
\Omega_n= \sum_r \Omega_{ r,n} (-1)^{cr}. 
\ee
We can then write
 \be
 \label{f10-final}
\ba
\mfF^{(1|0)}&= \left( {g_s \over \alpha} \right)^{c (c-2N)/2} {  (2 \pi)^{c/2-1} \over \prod_{k=0}^{c-1}\Gamma(N-k)}
 \re^{-{\CA_0 \over g_s}- {4 \pi^2 \ri n \over g_s}}
 \exp \left[ \widehat{\CF}^{\rm reg}((N-c)g_s; g_s) - \widehat{\CF}^{\rm reg}(N g_s ;g_s) \right]\\
 &\quad \times \Biggl\{\Omega_n(N) \Biggl[1 -{c\over 2} + c (N-c) \left(\log\left( {g_s\over \alpha} \right)  + \psi(N+1-c)-1 \right)\\
 &\qquad + {\CA_0+4\pi^2\ri n \over g_s}+ c g_s \partial_t \widehat{\CF}^{\rm reg}((N-c) g_s;g_s) \Biggr] -(N-c)\Omega_n'(N) \Biggr\}.
\ea
\ee
Here the hat denotes the additional subtraction of the linear and quadratic genus-zero terms from the regular free energy:
\be
\label{lin-quad}
\widehat{\CF}^{\rm reg}(t;g_s)=\CF^{\rm reg}(t;g_s)
-{1\over g_s^2}\left(f_{0,1} t -{t^2 \over 2} \log(\alpha)\right).
\ee
The unhatted $\CF^{\rm reg}$ retains its definition in (\ref{fg-cr}), and $\alpha$ is a numerical constant. The subtracted terms are incorporated explicitly in (\ref{f10-final}): the quadratic term gives the prefactor involving $\alpha$, while the linear term gives the instanton action
\be
\CA_0= c f_{0,1},
\ee
and is obtained by evaluating $c\partial_{t} \CF_0$ at the conifold point $t=0$. Finally, the last term in the expression (\ref{f10-final}) is 
\be
\Omega'_n(N)= -2 \pi \ri \sum_r r \Omega_{r,n} \re^{- 2 \pi \ri r N + \pi \ri r c}. 
\ee

Sometimes it is useful to work directly with the partition function, which we define here to be $\exp(\widehat{\CF}^{\rm reg})$. The one-instanton amplitude is then,  
\be
\label{z10-final}
\mathfrak{Z}^{(1|0)}=\exp\left(\widehat{\CF}^{\rm reg}(N g_s ;g_s) \right) \mfF^{(1|0)}.
\ee

Multi-instanton amplitudes for fixed $N$ are obtained in a similar way. We have to regularize higher order derivatives of the conifold 
free energy, and this is done as
\be
\label{higher-con}
g^k_s \partial^k_t \CF^{\rm con}(t-c g_s;g_s) \rightarrow \partial^k_N \log G_2(N+1-c) + \delta_{k2} \log(g_s), \qquad k\ge 2. 
\ee
After removing the common factor $\exp[(\CF_0^\CA(t)-\CF_0(t))/g_s^2]$ from the partition function, so that its perturbative sector uses the conventional prepotential $\CF_0$, the exponential factor $\exp(\CF_{[-\ell]})$ in (\ref{multi-ex}) is regularized as
\be
\begin{gathered}
(2 \pi)^{c \ell /2} {g_s^{c\ell(c\ell-2N)/2} \over \prod_{k=0}^{c\ell-1}\Gamma(N-k)}
\re^{-2\pi\ri r\ell N+\pi\ri rc\ell^2-{4\pi^2\ri n\ell\over g_s}}\\
\times \exp\left[\CF^{\rm reg}((N-c\ell)g_s;g_s)\right],
\end{gathered}
\ee
and the prefactors of this exponential, which involve higher derivatives of the free energy, are regularized by using (\ref{higher-con}). One has, for $k\ge 2$,  
\be
\hbar^k \CF^{(k)}_{-\ell} \rightarrow \hbar^k \partial_t^k \CF^{\rm reg} ((N-c \ell) g_s;g_s) + c^k\partial^k_N \log G_2(N+1-c\ell ) + \delta_{k2}\left(c^2\log(g_s)+2\pi\ri rc\right). 
\ee

\sectiono{Example 1: matrix models}
\label{sec-mm}

In the previous section we have shown how the general large $N$ results for instantons obtained in \cite{multi-multi,gm-multi,marmir, im} lead to 
results at finite $N$, after an appropriate regularization of the behavior at the conifold point. In this section we will test some of these results 
in the case of matrix models. 

\subsection{The one-cut cubic matrix model}

Let us first consider the one-cut matrix model. We will follow the conventions of \cite{marmir}, in which the potential is given by 
\be
V(x)=  {x^3 \over 3}-x. 
\ee
The coefficients $c_n(N)$ in (\ref{oc-pert}) can be computed explicitly by e.g. calculating the genus $g$ free energies $\CF_g(t)$ and re-expanding in $t$. One finds, for the very first 
orders, 
\be
\ba
c_1(N)&={N \over 48} (1+ 4 N^2), \\
c_2(N)&={N^2 \over192} (7+  8 N^2), \\
c_3(N)&= \frac{N}{9216} \left(336 N^4+664 N^2+105\right). 
\ea
\ee
We do not have an explicit expression for the coefficients $c_n(N)$ for arbitrary $n$, except in the replica limit,  
where we have found the expression\footnote{After the first version of this paper appeared in the arXiv, the formula (\ref{fg1-cubic}) was proved in \cite{laub-master} by using the techniques in Appendix 6 of \cite{biz}.} 
\be
\label{fg1-cubic}
f_{g,1}=2^{3-8g} 3^{-g} {(6g-4)!  
\over g! (3g-2)!}, \qquad g\ge 1. 
\ee

The asymptotic behavior of the series $c_n(N)$, for fixed $N$, should 
follow from the instanton amplitude (\ref{one-cut-ell}), (\ref{lone-ex}). The values of the functions and parameters appearing in those formulae can be read off from the results for the two-cut cubic matrix model in e.g. \cite{kmt,hk06,kmr,marmir}. One has 
\be
\label{Aoc}
\CA={4 \over3}
\ee
and one finds, for the very first functions $f_m^{(\ell)}(N)$, 
\be
 f_0^{(\ell)}(N)=-3 \log(2) \ell (N- \ell), \qquad 
 f_1^{(\ell)}(N)=\frac{\ell}{24}+\frac{17 \ell^3}{12}+\frac{7 \ell N^2}{8}-\frac{17 \ell^2 N}{8}.  
 \ee
 It is possible to verify numerically, for various values of $N$, that the one-instanton amplitude describes correctly the asymptotic 
 behavior of the sequence $c_n(N)$. 
 
A particularly simple case of the one-instanton amplitude is the replica limit, since it involves the perturbative free energies of the 
cubic matrix model in the so-called SW slice \cite{kmr, marmir} in which $t_1=-t_2$:
 \be
 \sum_{m \ge -1} f_m ^{(1)} (0) g_s^m=
-\sum_{g \ge 0} g_s^{2g-2}  \CG_g(-g_s, g_s), 
 \ee
 and one finds that the replica limit of the one-instanton amplitude is given by 
\be
\label{replica-oc}
\mathfrak{Z}^{(1)}_R (g_s)=   {g_s \over 8} \re^{-{4  \over 3 g_s}}\exp\left( -{35 g_s \over 24} + {35 g_s^2 \over  16}- {27685 g_s^3\over 4608}+ \cdots \right). 
\ee

It turns out that the prediction (\ref{replica-oc}) for the replica limit can be verified analytically. The reason is that the sequence 
(\ref{fg1-cubic}) is holonomic and its generating function 
\begin{equation}
\label{0fg1}
f(x)=\sum\limits_{g=1}^{\infty} f_{g,1} x^g
\end{equation}
satisfies an ODE which can be found with the package {\it Guess} due to Manuel Kauers (see e.g. \cite{kauers} for a survey of 
holonomic techniques). One finds, 
\begin{equation}
x^3 f^{(4)}(x)+\frac{11}{2} x^2 f^{(3)}(x)+\left(\frac{197 x}{36}-\frac{4}{9}\right)
   f''(x)+\frac{35 f'(x)}{72}=0. 
\end{equation}
It is easy to see that this ODE has a general trans-series solution of the form 
\begin{equation}
f(x) = f^{(0)}(x) + \CC_1 \text{e}^{-\frac{4}{3\sqrt{x}}}f^{(1)}(x)+ \CC_2 \text{e}^{\frac{4}{3\sqrt{x}}}f^{(-1)}(x),
\end{equation}
where $f^{(0)}(x)$ is the generating function (\ref{0fg1}), while $f^{(\pm 1)}(x)= x \Sigma(\pm x^{1/2})$, and $\Sigma(z)$ is given by 
\be
\Sigma(z)= 1-\frac{35 z}{24}+\frac{3745 z^2}{1152}-\frac{805805 z^3}{82944}+\frac{289554265 z^4}{7962624}-\frac{31241084875
   z^{5}}{191102976}+\CO\left(z^6\right). 
\ee
We have checked that this series agrees with the exponentiated series appearing in the one-instanton amplitude 
in (\ref{replica-oc}), after setting $z=g_s$, up to high order in $z$. The series $f^{(-1)}(x)$ corresponds to the replica limit of the ``negative" instanton amplitude $Z^{(0|1)}$. We note that the trans-series solution to the ODE gives only one-instanton amplitudes, in 
agreement with the fact that higher order instanton amplitudes
vanish in the replica limit.

\subsection{The two-cut cubic matrix model }

We will now consider the two-cut matrix model, in which there are two partial 't Hooft parameters $t_{1,2}$. The sum 
\be
-{\alpha \over 4}= t_1+ t_2
\ee
 is a mass parameter, and we will use as flat coordinate $t=t_1$. 
 It was noted in \cite{kmr,marmir} that the theory simplifies on the ``slice" $\alpha=0$, to which we will restrict from now on. 
 The free energies of this theory have a conifold 
behavior of the form 
\be
\label{alpha-gap}
\CF_g(t) ={B_{2g} \over g(2g-2)} \frac{1}{t^{2g-2}}+ \CO(t), \qquad g\ge 2, 
\ee
which is twice the usual one. For the leading instanton effect, \cite{marmir} found $c=1$, 
which we will use from now on. The calculation of the regulated 
amplitude is as above, and we just have to include a factor of $2$. We obtain 
\be
\ba
\mathfrak{F}^{(1|0)}(g_s;N)&= {g_s^{1-2N} \over \Gamma^2(N)}
 \left[ 2(N-1) \left(-1+ \psi(N)\right) + 2 \log(g_s) (N-1) + g_s \partial_t \CF^{\rm reg}((N-1) g_s;g_s) \right] \\
&\qquad  \times \exp \left[ \CF^{\rm reg}((N-1)g_s; g_s) - \CF^{\rm reg}(N g_s ;g_s) \right].
\ea
\ee
This can be rewritten explicitly as a series in $g_s$, as follows
  \be
  \ba
\mathfrak{F}^{(1|0)}(g_s;N) &= \left( { g_s \over 8}  \right)^{1-2N}{1 \over \Gamma^2(N)}  \re^{-{\CA \over g_s}}\exp \left( -\sum_{m \ge 1} M_m(N) g_s^m \right) \\ & \qquad \times 
\left( {\CA \over g_s} + 2 \log(g_s) (N-1) +\sum_{p \ge 0} d_p(N) g_s^p \right), 
\ea
\ee
where $\CA$ has the same value as in the one-cut case (\ref{Aoc}), and the very first coefficients appearing in this expression can be computed to be
\be
\ba
d_0(N)&= 2(N-1) \left( -1+ \psi(N)- 3 \log(2) \right), \\
d_1(N)&={17 \over 4}(N-1)^2 + {1\over 24},\\
M_1(N)&={35 \over 24} + {17 \over 4} N(N-1). 
\ea
\ee
In order to determine the leading asymptotics of the coefficients in the perturbative series (\ref{oc-pert}) we have 
to consider as well the anti-instanton amplitude. The calculation is similar to the one in the one-cut case. Roughly, 
it is obtained from $\mathfrak{F}^{(1|0)}$ by switching $N \rightarrow -N$, $g_s \rightarrow -g_s$, and one finds, 
  \be
  \ba
\mathfrak{F}^{(0|1)} (g_s;N)&= (-1)^{2N} \left( { g_s \over 8}  \right)^{1+2N}{1 \over \Gamma^2(-N)}  \re^{{\CA \over g_s}}\exp \left( -\sum_{m \ge 1} M_m(-N) (-g_s)^m \right) \\ & \times \left( -{\CA \over g_s} + 2 \log(g_s) (-N-1) +\sum_{p \ge 0} d_p(-N) (-g_s)^p \right).  
\ea
\ee
To write the explicit formula for the asymptotics, we note that the instanton amplitudes involve a logarithmic term $\log(g_s)$, 
therefore one has to 
use the more complicated large order formulae of e.g. \cite{gikm}. Let us write the amplitude as 
 \be
 \label{10series}
 \mathfrak{F}^{(1|0)}(g_s;N)=  \re^{-{\CA \over g_s}} g_s^{-b-1} \sum_{n \ge 0} \mu_n^{(1)} (N)g_s^n+ \re^{-{\CA \over g_s}} \log(g_s) \, g_s^{-b} \sum_{n \ge 0} \mu_n^{(2)}(N) g_s^n, 
 \ee
 and let us define 
  \be
  \label{asym-10}
  \ba
f_k(N)= & {1\over 2 \pi} \CA^{-k-b-1}\Gamma(k+b+1) \sum_{n \ge 0} { \mu^{(1)}_n(N) \CA^n \over \prod_{i=1}^n (k+b+1-i)} \\
 &+ {1\over 2 \pi}  \CA^{-k-b}\Gamma(k+b) \sum_{n \ge 0} { \mu^{(2)}_n (N) \CA^n \over \prod_{i=1}^n (k+b-i)} \left(\log(\CA)- \psi(b+k-n) \right), 
 \ea
 \ee
 where 
 \be
 b= 2N-1. 
 \ee
We note that $\psi(k+b)$ should be understood as an asymptotics series in $1/k$, 
\be
\psi(k+b) \sim \log(k) + {2b-1 \over 2k}+ \CO(k^{-2}). 
\ee
Then, the asymptotic formula is 
\be
\label{ck-as}
c_k(N) \sim f_k(N)+ (-1)^k f_k (-N), \qquad k \gg 1. 
\ee
When $N$ is an integer, only one of the two terms in the r.h.s. of (\ref{ck-as}) contributes. This asymptotic formula can be verified numerically for various values of $N$. 
%
%
%
 %
%
%
%
%
%
%

\sectiono{Example 2: Seiberg--Witten theory}
\label{sec-sw}

In this section we will analyze in detail the case of topological strings on the SW curve. This is one of the simplest 
examples of topological string models and it will illustrate many aspects of the theory. In particular, we will give 
evidence that the resurgent structure of the topological string free energies has information on the BPS spectrum 
of the theory, and changes discontinuously as we cross the curve of marginal stability. Then we will analyze the conifold 
free energies at fixed $N$, which for integer $N$ are described by the Bonelli--Grassi--Tanzini (BGT) matrix model 
\cite{bgt}.  

\subsection{The topological string free energies}

The prepotential of $\CN=2$ supersymmetric Yang--Mills theory was derived by Seiberg and 
Witten in their celebrated work \cite{sw}. Later on, it was shown that it can 
be obtained by ``geometric engineering," namely it can be 
obtained by a certain four-dimensional limit of topological string theory on a non-compact 
CY manifold \cite{kkv} (for reviews of SW theory and its stringy 
origin, see e.g. \cite{lerche,klemm-sw}). The prepotential corresponds to 
the genus zero free energy of the topological string. One advantage of this perspective is that it provides an 
infinite set of amplitudes in SW theory, corresponding to the four-dimensional limit of the higher genus 
topological string free energies. 

A more abstract formulation of this sequence of free energies is inspired by the B-model approach of 
\cite{bkmp}. In this formulation, the free energies are associated directly to the SW curve
\be
\label{SWcurve}
y^2+2 \cosh(x)= 2 u.  
\ee
These free energies can be obtained by using topological recursion \cite{eo} or, more conveniently in our case, 
by solving the holomorphic anomaly equations (HAE) of 
\cite{bcov} on the curve, as first pointed out in \cite{hk06}. Let us now summarize how to solve the HAE 
in this case. We will follow the ``holomorphic" formulation used in \cite{gm-multi}, rather than the ``modular form" formulation used in \cite{hk06}. 

The complex variable $u$ in (\ref{SWcurve}) parametrizes the modulus of the curve and will be our complex modulus. The 
periods of the curve are given by 
\be
\ba
a(u)&={2 {\sqrt{2}} \over \pi} {\sqrt{u+1}} E\left( {2 \over u+1} \right),  \\
a_D(u)&={\ri  \over 2} (u-1)  {}_2 F_1 \left( {1\over 2}, {1\over 2}, 2;{ 1-u \over 2} \right),
\ea
\ee
where we follow the conventions in e.g. \cite{fb}\footnote{In particular, our convention for $a$ is different from the one in \cite{hk06}: our $a$ is their $2a$.}. As usual, the periods provide flat coordinates for the moduli space. From the point of view of SW theory, they give the central charges of BPS states in the spectrum. Let $\gamma=(\gamma_e, \gamma_m)$ be a charge, where $\gamma_{e,m}$ refers to the electric (respectively, magnetic) charge. Then, the central charge is given by 
\be
Z_\gamma (u)= \gamma_e a(u) + \gamma_m a_D(u). 
\ee
The period $a(u)$ is appropriate as a flat coordinate for the large radius region $u \rightarrow \infty$, which in 
SW theory corresponds to the semiclassical region where the $\CN=2$ theory is 
weakly coupled. There are conifold points at $u=\pm 1$. At $u=1$, known as the monopole point, we have $a_D(u=1)=0$, and 
a magnetic monopole in the spectrum of the theory becomes massless. Near $u=1$ we will use the local coordinate
\be
v=u-1,
\ee
which vanishes at the monopole point. At the other conifold point at $u=-1$, one has
\be
a_D(-1)+  a(-1)=0, 
\ee
and a dyon in the spectrum with charge $\gamma=(1,1)$ becomes massless. 

In order to construct the topological string free energies it will be convenient to use the flat coordinate near the monopole point given by  
\be
\label{taD}
t=- \ri a_D.  
\ee
We define the (magnetic) prepotential as 
\be
\label{dualF}
{\partial \CF_0 \over \partial t} =- \pi a, 
\ee
which has the local expansion 
\be
\CF_0(t)= {t^2 \over 2} \left( \log\left( {t \over 16} \right)-{3 \over 2} \right) - 4 t - {t^3 \over 16}+ \cdots
\ee
In this section, the calligraphic free energies $\CF_g(t)$ denote holomorphic free energies, while
Roman free energies $F_g$ denote their non-holomorphic counterparts obtained by using the HAE.
We will mostly work in the frame appropriate for the conifold or monopole point at $u=1$, and we will sometimes refer 
to it as the ``magnetic" frame. The frame associated to the flat coordinate $a(u)$ will be called ``electric" frame. In the following, we will work in the magnetic frame, unless stated otherwise. 

 To set up the HAE we need various ingredients (we refer to \cite{hk06,gm-multi} for the relevant background). The first one is the Yukawa coupling, 
which is given by 
\be
C_u ={1\over 2(u^2-1)}. 
\ee
The second ingredient is the genus one free energy, which is given in the magnetic frame by
\be
\CF_1=-{1\over 2} \log\left( -{\rd t \over \rd u} \right)- {1\over 12} \log(16(u^2-1)). 
\ee
%
%
%
It has the following local expansion, 
\be
\CF_1= -{1\over 12} \log(t) +{t \over 32}-{3 t^2 \over 512}+{19 t^3 \over 12288}+ \CO(t^4). 
\ee
The genus one free energy determines the propagator $S$ to be used in the HAE. 
We will follow the conventions in \cite{gm-multi} for the propagator:
\be
S= {2 \over C_u} \partial_u F_1. 
\ee
This is also the propagator used in \cite{hk06}. Its holomorphic limit can be written as \cite{gm-multi}
\be
\CS= -{1\over C_u} {T''(u) \over T'(u)} - \mathfrak{s}(u), 
\ee
where $T(u)$ is a generic flat coordinate, and $\mathfrak{s}(u)$ depends only on the modulus $u$ and is given by 
\be
\mathfrak{s}(u)= {2u \over 3}. 
\ee
The propagator satisfies the equation 
\be
\label{ds}
\partial_u S\equiv S^{(2)}=C_u \left(S^2 + 2 \mathfrak{s}(u) S +  \mathfrak{f}(u)\right), 
\ee
and in our case
\be
\mathfrak{f}(u)= {u^2 \over 9} + {1\over 3}. 
\ee
We can write the following explicit expression in the magnetic frame, 
\be
\CS(v)= {1 \over 6 v C_u} \left\{ -\frac{3 E\left(\frac{v}{v+2}\right)}{K\left(\frac{v}{v+2}\right)}+ {4 +v \over 2+v} \right\},  
\ee
where $K$, $E$ are elliptic integrals of the first and second kind, respectively, and their argument is the modulus square. 
There is a similar expression in the electric frame, 
\be
\CS^{\rm E}(u)={1\over 2 C_u (u-1)}  {E \left( {2 \over 1+u} \right) \over K \left( {2 \over 1+u} \right)}- {2 u \over 3}. 
\ee

In the direct integration approach to the HAE, the free energies are viewed as functions of $S$ and the modulus $u$. To formulate 
these equations, we introduce a derivation $\mD_u$ which acts on a function of $S$ and $u$ as follows:
\be
\mD_u f(S,u)= \partial_S f(S, u) S^{(2)} + \partial_u f(S,u), 
\ee
where $S^{(2)}$ was defined in (\ref{ds}). We introduce as well the connection associated 
to the K\"ahler metric on the moduli space, through its Christoffel symbol 
\be
\label{chris-local}
\Gamma^u_{uu}= -C_u \left( S+ \mathfrak{s}(u) \right).
\ee
We have all the ingredients to write down the HAE:
\be
\label{f-hae}
{\partial F_g \over \partial S}= {1\over 2} \left( {\cal{D}}_u \mD_u F_{g-1}+
 \sum_{m=1}^{g-1} \mathfrak{D}_u F_m \mD_u F_{g-m} \right), \qquad g\ge 2. 
\ee
Here, ${\cal{D}}_u$ is the covariant derivative w.r.t. the K\"ahler metric, and it acts on the indexed object $\mD_u F_g$ as
\be
{\cal{D}}_u \mD_u F_{g}=\mD_u^2 F_g -\Gamma^u_{uu} \mD_u F_{g}. 
\ee
One can now use as an input the genus one free energy, and solve for $F_g$ recursively. For example, for $g=2$ one finds, 
\cite{gm-multi}
\be
\label{f2-local}
F_2(S,u) = C_u^2 \left\{  {5 S^3 \over 24} +{S^2 \over 24} \left( 9 \mathfrak{s}(u) 
+ 3 {C'_u \over C_u^2} \right) + {1\over 4} S \mathfrak{f}(u) \right\} +f_2(u). 
\ee
In this equation, $f_2(u)$ is an arbitrary holomorphic function of $u$ which is not fixed by the HAE. It is called the holomorphic ambiguity and it occurs at every genus $g$. Additional information is required to fix it. As 
already pointed out in \cite{bcov}, one needs a reasonable ansatz which takes into account the behavior at special points in moduli space. In the case of SW theory, 
this ansatz was found in \cite{hk06}, and takes the form, 
\be
f_g(u)= u^{3-g} \sum_{i=1}^{2g-2} {a^{(g)}_{i} \over (u^2-1)^i}. 
\ee
At each genus $g$ there are $2g-2$ coefficients $a^{(g)}_{i} $ 
that need to be determined. As pointed out in \cite{hk06}, one can impose the 
gap behavior at the conifold, (\ref{fg-cr}). 
This gives $2g-2$ conditions, since it prescribes the values of the coefficients of $t^{-1}, \cdots, t^{-(2g-2)}$ 
in the expansion, therefore it fixes completely the ambiguity. 

The gap behavior would arise naturally if the free energies $\CF_g(t)$ 
were obtained from the large $N$ expansion of a matrix model. Indeed, 
one of the consequences of the TS/ST correspondence of \cite{ghm} is that, for local CY manifolds, the 
topological string free energies in the conifold frame are given 
by a matrix model which can be obtained from the quantization of the curve, as explained in 
\cite{mz}. In the case of SW theory, this matrix model was written down explicitly in 
\cite{bgt} and we will refer to it as the BGT matrix model. It was previously studied in \cite{gm}, 
where it was called the polymer matrix model. It is defined by 
the partition function
\be
\label{bgt}
\mZ(N; g_s)= {1\over N!} \int \prod_{i=1}^N {\rd x_i \over 4 \pi} \prod_{i<j} \tanh^2 \left( {x_i - x_j \over 2} \right) \, \exp \left[ - {4 \over g_s} \sum_{i=1}^N \cosh(x_i) \right]. 
\ee
The flat coordinate near the monopole point (\ref{taD}) is identified with the 't Hooft parameter of the matrix model,  
\be
t = N g_s, 
\ee
and the $1/N$ expansion of the free energy $\mF(N;g_s)= \log\, \mZ(N; g_s)$ gives the genus $g$ free energies of SW theory 
in the magnetic frame, 
\be
\mF(N; g_s) \sim \sum_{g\ge 0} \CF_g (t) g_s^{2g-2}. 
\ee

\subsection{Resurgent structure and BPS spectrum}

Recently there has been growing evidence that the resurgent structure of the topological string on arbitrary CY threefolds contains 
information about BPS invariants \cite{mm-s2019,gm-peacock,astt, gm-multi, ghn,aht,gu-res,gkkm}. A precise and general conjecture can be found in \cite{im} (see \cite{amp} for the refined case, and \cite{ms} for the extension to the real topological string). It states the following:

\begin{enumerate} 

\item The total topological string free energy in a given frame is a resurgent function. Its Borel singularities are integer linear combinations of the CY periods. 

\item The singularities display a multi-covering structure: given a singularity $\CA$, all its 
integer multiples $\ell\CA$, $\ell \in \IZ\backslash \{0\}$, appear as singularities as well. The Stokes automorphism 
for the singularities occurring along a half-ray $\ell \CA$, $\ell \in \IZ_{>0}$ is given by (\ref{stokes-afr}) (for the case $c=0$) 
or (\ref{stokesZ}) (for the case $c\not=0$).

\item The Stokes constant $\Omega$ appearing in the Stokes automorphism through (\ref{aS}) is the BPS invariant associated to the 
charge $\CA$.  

\end{enumerate}

We note that the possible Borel singularities and their Stokes constants do not depend on the frame. What depends on the frame is the 
form of the multi-instanton amplitudes. This is further clarified in Appendix \ref{appendix-a}. 

\begin{figure}
  \centering
  \includegraphics[height=6cm]{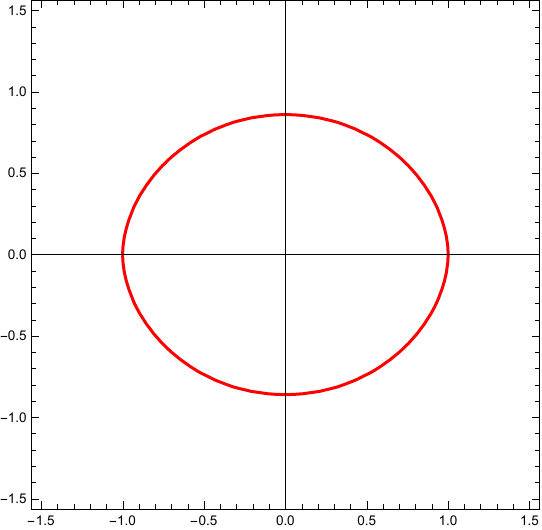} 
  \caption{The curve of marginal stability in the $u$-plane, defined by the equation (\ref{cms-eq}). }  
  \label{cms-fig}
\end{figure}

Based on the above conjectures, we expect that the resurgent structure of the 
sequence $\CF_g(t)$ in the magnetic frame encodes the 
BPS spectrum of SW theory. This spectrum has been studied intensively since the original work \cite{sw}. The main results 
are the following \cite{fb}. First of all, the spectrum depends on the value of the modulus $u$. 
Inside the curve of marginal stability, defined by 
\be
\label{cms-eq}
{\rm Im} \left( {a_D \over a} \right)=0, 
\ee
we have the so-called strong coupling spectrum: the only stable states are the monopole with charge $\gamma_M= (0,1)$ and the dyon with charge $\gamma_D=(1,1)$. Outside 
this curve, we have the so-called weak coupling spectrum of states, consisting of a $W$ boson and a tower of dyons, 
with charges, respectively, 
\be
\gamma_W=(1,0), \qquad \gamma_n=(n, 1), \quad n \in \IZ. 
\ee
The states with opposite charges also belong to the spectrum. In \figref{cms-fig} we show the curve of marginal stability in the $u$-plane. 

The study of the SW spectrum has triggered many interesting developments. In \cite{klmvw} a string-inspired argument 
led to a determination of the spectrum based on geodesics on the SW curve. On the other hand, the quantization of the 
SW curve leads to the Schr\"odinger operator of the modified Mathieu equation, and by using the all-orders WKB method one 
can upgrade the classical periods $a(u)$, $a_D(u)$ to quantum periods, which are formal power series in 
$\hbar^2$ \cite{mirmor,ns}. Although this was not appreciated at the time, 
the criterion of \cite{klmvw} to find stable BPS states is exactly the same criterion 
that determines the position of Borel singularities of the  
quantum periods associated to the SW curve \cite{dpham,reshyper}. 
A direct, numerical verification of the connection between these Borel singularities 
and the BPS spectrum was carried out in \cite{ggm}. 
The connection between BPS states and the WKB method implicit in \cite{klmvw} was developed in \cite{gmn}, which formulated 
a powerful method to obtain BPS spectra in $\CN=2$ supersymmetric field theories, and made contact with the results of \cite{ks}.  

According to the conjectural connection between BPS spectra and topological strings, the resurgent structure 
of the topological string on the SW curve should give a new way of obtaining the SW spectrum. In particular, the 
Borel singularities of the series of genus $g$ free energies should reflect the spectrum of stable BPS states, for a given value of $u$, and the 
Stokes discontinuities should encode information on the corresponding BPS invariants.  More precisely, the possible location of primitive Borel 
singularities is given by 
\be
\CA_\gamma= 2 \pi \left( \gamma_e a + \gamma_m a_D \right), 
\ee
where $\gamma$ is the charge of the corresponding BPS state. As we explained before, 
a single BPS state of charge 
$\gamma$ leads to an infinite tower of singularities $\pm \ell \CA_\gamma$,  $\ell \in \IZ_{>0}$. 
In particular, if the conjectural connection 
between resurgent structures and BPS spectra is correct, we should see a 
different set of Borel singularities depending on whether 
we are inside or outside the curve of marginal stability. 
 
\begin{figure}
  \centering
  \includegraphics[height=3.7cm]{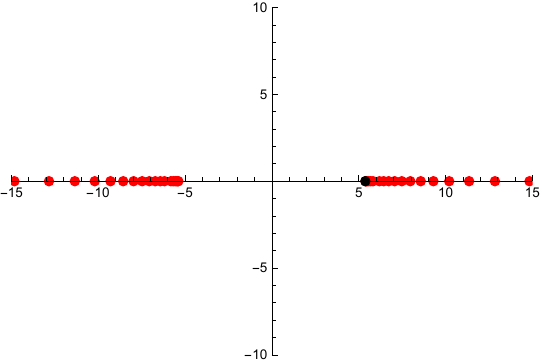}\qquad \qquad  \includegraphics[height=3.5cm]{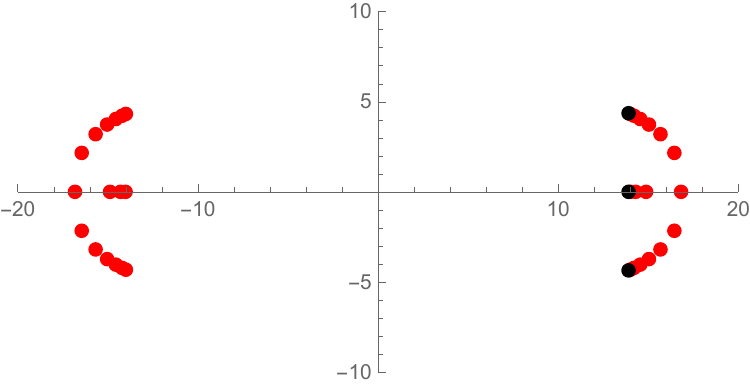} 
  \caption{The Borel singularities of $\CF^{\rm reg}(t)$ in SW theory, inside the curve of marginal stability with $u=1/2$ (left), and outside the curve with $u=5/2$ (right). The black dot in the figure in the left signals the value $2 \pi (a + a_D)$, and the black dots in the figure in the right signal the value $2 \pi a$ (on the positive real axis), $2 \pi (a+a_D)$ (above the positive real axis) and $2 \pi (a-a_D)$ (below the positive real axis).}  
  \label{cross-fig}
\end{figure}

 It turns out that all these expectations can be verified numerically in detail, confirming the 
 conjectural connection in this particular case. Let us consider the sequence of 
 magnetic free energies $\CF_g(t)$. The conifold behavior (\ref{fg-con}) implies 
 the existence of a tower of singularities at 
\be
\pm \ell \CA_{0,1}, \qquad \CA_{0,1}= 2 \pi \ri t, \quad \ell \in \IZ_{>0}.  
\ee
There should be in addition a tower of singularities at
\be
\pm \ell \CA_{1,1}, \qquad \CA_{1,1}= -2 \pi a - 2 \pi \ri t,
\ee
due to the dyon. To see these dyon singularities it is convenient to subtract the pole in $t$ and to 
consider the regularized free energies $\CF^{\rm reg}_g(t)$ defined in (\ref{fg-cr}). In 
this way, we subtract the leading singularity in $\CF_g(t)$ 
corresponding to the monopole. In the regularized free energies, the dyon singularities are the dominant ones, therefore 
easier to identify numerically. One finds indeed that, inside the curve of marginal stability, these free energies 
have a singularity at $\pm \CA_{1,1}$, due to the dyon, see the plot on the left of \figref{cross-fig}. For real $u$, 
these are on the real axis. 

As we cross the curve of marginal stability, the singularity at $ \CA_{1,1}$ splits into two singularities $(\pm 1, 1)$, 
and the W-boson with charge $(1,0)$ (a similar splitting occurs of course for the singularity at $-\CA_{1,1}$). This is 
clearly seen in the \figref{cross-fig}, on the right, 
and it agrees with the well-known wall-crossing behavior of SW theory. In principle, one should find in the Borel plane 
singularities corresponding to the tower of dyons $\pm (n, 1)$, with $|n|>1$. However, numerical approximations to the singularity 
structure, like the one depicted in \figref{cross-fig}, are based on a finite number of terms in the factorially 
divergent sequence (in this case, we are using $100$ terms in the sequence of free energies). Due to their approximate nature, 
they display only the singularities 
which are closer to the origin. 

In addition to checking the presence of Borel singularities we can compute numerically the corresponding 
Stokes constants. When 
the corresponding singularities are on the 
real axis, the most efficient way to extract these constants is to study the large order behavior of the sequence $\CF_g(t)$, which is determined at leading 
order by the one-instanton amplitude (\ref{ntinst}). In doing this calculation for the dyon singularity $\CA_{1,1}$, it is useful to note that
\be
\CA_{1,1} = 2 {\partial \CF_0 \over \partial t}- 2 \pi \ri t, 
\ee
therefore $c=2$ in this case. This defines a modified prepotential 
\be
\widetilde \CF_0 = \CF_0 -{\pi \ri t^2 \over 2}. 
\ee
For the singularities which are not in the real axis, the best way to obtain a numerical approximation to their Stokes constant is the following. Given a finite number of terms in the perturbative series, one calculates the analytic continuation of its Borel transform by using a Pad\'e approximant. What is plotted in \figref{cross-fig} is precisely the poles of this approximant. The line of poles coming out of each singularity mimics the branch cut starting there. The discontinuity across this branch cut can be approximated numerically by summing the residues of the poles of the approximant. Comparing this discontinuity to the instanton formula, we extract the Stokes constant. The numerical study shows that the Stokes constants $\Omega_\gamma$ for the singularities in the strong coupling region are 
\be
\Omega_{(0,1)}= \Omega_{(1,1)}=1,
\ee
while in the weak coupling region we have
\be
\Omega_{(1,1)}=1, \qquad  \Omega_{(1,0)}=-2, 
\ee
in agreement with standard results.

\subsection{Perturbative series}

The analysis in the previous section was made in the large $N$ theory. As we have done in the case of Hermitian 
matrix models, we can consider the subtracted regular SW free energy in the
magnetic frame $\widehat{\CF}^{\rm reg}(t;g_s)$, set $t= Ng_s$, and re-expand in $g_s$ to
obtain the conventional perturbative series. We find in this way, 
\be
\label{fNSW}
\mathfrak{F} (g_s; N)= {1\over 32} N(1-2 N^2) g_s + {1\over 512} N^2 (5 N^2-3) g_s^2 +\frac{N \left(-66 N^4+38 N^2+3 \right) }{24576}  g_s^3+ \CO(g_s^4). 
\ee
Note that in calculating $\mathfrak{F} (g_s; N)$ we have omitted from $\CF_0$ the linear 
and quadratic terms, which in this case read
\be
-4 t -{t^2 \over 2} \log(16). 
\ee

We would like to determine the resurgent structure of the above series in $g_s$, and in particular relate it to the 
general resurgent structure at large $N$. The important new ingredient that we find in this case is the following. When we set 
$t= N g_s$ and we expand around $g_s =0$, we are effectively zooming on the monopole point. 
However, this is a point at which the resurgent structure of the large $N$ theory is not well-defined, strictly speaking, since it 
is located at the curve of marginal stability. At the same time, finding the resurgent structure of the series (\ref{fNSW}) 
is a well-defined problem with a unique answer for each value of $N$. As we will explain in the following, our 
results indicate that the resurgent structure of (\ref{fNSW}) depends on $N$ in a continuous way, inherited 
from the {\it strong coupling spectrum} of the underlying SW theory. In particular, we have not found evidence of wall--crossing phenomena 
in the perturbative theory at finite $N$.

To address these issues, we consider the one-instanton partition function 
(\ref{z10-final}). In the SW case, $c=2$, $f_{0,1}=-4$, 
and $\alpha=16$. There are no singularities with $n\not=0$ in (\ref{acr}), therefore we have
 \be
 \label{z10SW}
 \ba
 \mathfrak{Z}^{(1|0)}(g_s;N)&= \left( {g_s \over 16} \right) ^{ 2(1-N)} {2 \over \Gamma(N) \Gamma(N-1)} \re^{8/g_s} \exp \left[ \widehat{\CF}^{\rm reg}((N-2)g_s; g_s) \right] \\
 &\quad \times \Biggl\{\Omega(N) \Biggl[(N-2) \left(\log\left({g_s \over 16} \right)+ \psi(N-1)-1 \right)\\
 &\qquad -{4 \over g_s}+ g_s \partial_t \widehat{\CF}^{\rm reg}((N-2) g_s;g_s) \Biggr] -{N-2\over 2}\Omega'(N) \Biggr\}.
 \ea
\ee
The hat denotes the subtraction defined in (\ref{lin-quad}), as in (\ref{f10-final}). The value of the instanton action is given by $ -2 \pi a(1)= -8$, and  
\be
\label{omegaN-sw}
\Omega(N)= \sum_r \Omega_r \re^{-2 \pi \ri r N}
\ee
is a sum over all the singularities which collapse to $\CA=-8$ at the conifold limit. As we explained above, 
since we are exactly at the curve of marginal stability, and the structure of singularities changes discontinuously at that 
point, it is not {\it a priori} clear which spectrum is appropriate in (\ref{omegaN}). If we consider the strong coupling spectrum, 
we have a single singularity at $-2 \pi (a+ a_D)$, i.e. $r=-1$ and 
\be
\label{omstrong}
\Omega_{\rm strong} (N) =  \re^{2\pi \ri N}.
\ee
If we consider the weak coupling spectrum, we have Borel singularities due to dyons at $-2 \pi (a\pm a_D)$, with 
$r=\pm 1$, and a W-boson singularity at $-2 \pi a$, with $r=0$. This gives
\be
\label{omweak}
\Omega_{\rm weak}(N)= 2 \left(\cos(2 \pi N)-1\right).  
\ee
Both $\Omega_{\rm weak}(N)$ and $\Omega_{\rm weak}'(N)=-4\pi\sin(2\pi N)$ vanish when $N$ is an integer,
so the weak coupling spectrum cannot give the right result for the one-instanton amplitude. As we will see, all our explicit
calculations confirm that the relevant spectrum to understand the resurgent structure of the perturbative series (\ref{fNSW}) is 
the strong coupling one, and one has to use $\Omega_{\rm strong} (N)$ in (\ref{z10SW}). 

To find further evidence for this, let us consider the resurgent structure of the 
perturbative series (\ref{fNSW}) for $N=1,2$, and compare it with the 
prediction of the one-instanton amplitude (\ref{z10SW}). Let us note that for $N=1$, the inverse Gamma functions in (\ref{z10SW}) 
give a zero, while the digamma function $\psi(N-1)$ gives a pole. The amplitude has to be evaluated as a limit, which is well-defined 
and leads to a simplified expression. 
For $N=1$ the resulting amplitude reads 
\be
\label{trans-N1-SW}
\mathfrak{Z}^{(1|0)}(N=1)= 2\, \Omega\,  \re^{8/g_s} \exp \left[ \mfF(-g_s; 1) \right],
\ee
while for $N=2$ it is given by 
\be
\label{trans-N2}
\mathfrak{Z}^{(1|0)}(N=2)= {512  \, \Omega \over g_s}\re^{8/g_s}\left(-{4 \over g_s^2}+ \partial_t \widehat{\CF}^{\rm reg}(0;g_s) \right) ={512 \, \Omega \over  g_s^3} \re^{8/g_s} \sum_{g \ge 0} f_{g, 1} g_s^{2g}. 
\ee
In these equations $\Omega=\Omega_{\rm strong}(1)= \Omega_{\rm strong}(2)=1$, as it follows from (\ref{omstrong}). The coefficients 
$f_{g,1}$ in (\ref{trans-N2}) are defined by 
\be
\CF_g^{\rm reg}(t)= f_{g,1} t+ \CO(t^2)
\ee
and they give the replica limit of the perturbative series (\ref{r-oc}). 
 
 It turns out that the expressions above for the one-instanton amplitudes can be verified analytically. This is because the BGT matrix 
 model (\ref{bgt}) provides explicit non-perturbative resummations of the perturbative series (\ref{fNSW}) for integer $N$. In particular, 
for $N=1,2$ one has \cite{bgt2}:
\be
\ba
\mZ(1;g_s)&={1\over 2 \pi} K_0 \left( {4 \over g_s} \right), \\
\mZ(2;g_s)&={1\over 32 \pi^{3/2}} G_{1,3}^{3,0}\left({16 \over g_s^2} \left|
\begin{array}{c}
 {3 \over 2} \\
 0,0,0
\end{array}
\right.\right).
\ea
\ee
Let us first consider the case $N=1$. The asymptotic expansion of $\mZ(1;g_s)$ is given by 
\be
\mZ(1;g_s)\sim  {g_s^{1/2} \over 4 {\sqrt{2\pi}}}  \re^{-4 /g_s} \varphi_K(g_s),
\ee
where
\be
\varphi_K (g_s)= \exp\left(\mfF(g_s;1) \right)= \sum_{n \ge 0} d_n g_s^n=1 -{g_s \over 32} + {9 g_s^2 \over 2048}+ \cdots
\ee
The resurgent structure of this series is the one of the special 
function $K_0(4/g_s)$, which was studied in \cite{mmbook}, Example 3.5. The 
corresponding trans-series or one-instanton amplitude is given by the asymptotic 
expansion of $ I_0(4/g_s)\sim K_0(-4/g_s)$, which gives the 
term $\exp(\mfF(-g_s;1))$ in (\ref{trans-N1-SW}). We also note that  
\be
d_n \sim {1\over \pi} (-8)^{-n} \Gamma(n), 
\ee
which is exactly what follows from (\ref{trans-N1-SW}) with $\Omega=1$. 

The case $N=2$, involving the Meijer function, is richer. We introduce the more convenient variable 
\be
y= {g_s \over 4}. 
\ee
The asymptotic expansion is 
\be
{1\over 32 \pi^{3/2}} G_{1,3}^{3,0}\left({1\over y^2} \left|
\begin{array}{c}
 {3 \over 2} \\
 0,0,0
\end{array}
\right.\right) \sim {\re^{-{2 /y}} \over 32 \pi  } y^2 \varphi_0(y), 
\ee
where
\be
\varphi_0(y)= \exp\left( \mfF(4 y; 2) \right)=\sum_{n \ge 0} b_n y^n=1-{7 y \over 4 }+ {117 y^2 \over 32 } + \cdots
\ee
By using the ODE which characterizes the 
Meijer function, one finds that $\varphi_0(y)$ satisfies the ODE
\be
\label{meijer-ode}
y^4 f^{(3)}(y)+(9 y+6) y^2 f''(y)+(y (19 y+24)+8) f'(y)+2 (4 y+7) f(y)=0. 
\ee
Therefore, we should be able to obtain the trans-series (\ref{trans-N2}) from a trans-series solution to this ODE. We then search for 
solutions of the form 
\be
f(y) = \re^{A/y} y^{-b_A} \varphi_{A}(y), \qquad \varphi_{A}(y)= 1+ \CO(y). 
\ee
We find that the possible values of $A$ are $A=2,4$, while $b_A$ are $3,0$, respectively. The 
solutions are related by $\varphi_4(y)= \varphi_0(-y)$, so that $\varphi_4(y)$ corresponds to the $(2|0)$ or two-instanton 
sector. The series $\varphi_2(y)$ has the form 
\be
\varphi_2 (y)= \sum_{n \ge 0} a_n y^{2n}, 
\ee
where the $a_n$ satisfy the recursion 
\be
\label{anrec}
a_{n+1}= {(2n-1)^3 \over 8 (n+1)} a_n, \qquad n\ge 0.  
\ee
If (\ref{trans-N2}) gives the correct trans-series, $\varphi_2(y)$ should be equal, up to an overall constant, to 
$g_s^2 \partial_t \CF^{\rm reg}(0;g_s)$. By using 
$f_{0,1}=-4$ and the recursion (\ref{anrec}), one finds the closed form expression 
\be
\label{fg1}
f_{g,1}=\frac{((2 g-3)\text{!!})^3}{2^{7 g-2} \, g!}, \qquad g\ge 0, 
\ee
which indeed agrees with the result of the HAE\footnote{The expression (\ref{fg1}) was conjectured by A. Klemm {\it circa} 2006 in unpublished work, based on 
the solution of the HAE to high genus.}. The overall coefficient of the trans-series (\ref{trans-N2}) can be extracted from the asymptotics of the coefficients $b_n$, which 
can be found to be 
 \be
 {b_n  \over 4^n} \sim {16^2  \over \pi} \left( -8 \right)^{-n-3} \Gamma(n+3) \sum_{g \ge 0} 
 { f_{g,1} (-8)^{2g} \over \prod_{i=1}^{2g} (n+3-i)}.  
 \ee
This is precisely as predicted by (\ref{trans-N2}) and confirms that $\Omega=1$. Finally, let us note that the above analysis indicates that the one-instanton amplitude
controlling the asymptotics of the replica limit (\ref{r-oc}) is essentially given by $\varphi_0(-y)$. This also 
agrees with the prediction from (\ref{z10SW}). 

\begin{figure}
\centering
\includegraphics[width=0.55\textwidth]{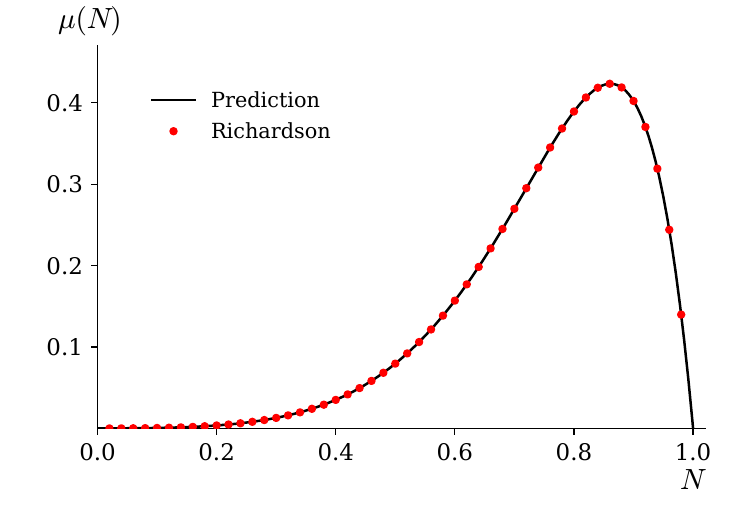}
\caption{The continuous curve is the prediction (\ref{fn-ex}) for $\mu(N)$. The dots are
obtained from the free-energy coeﬃcients (\ref{fNSW}), at $N= j/50$ with $j = 1,...,49$, by generalized
Richardson extrapolation.}
\label{fncomp-fig}
\end{figure}

In addition to the analytic study of the resurgent structure of the perturbative series when $N=1,2$, one can also consider 
arbitrary values of $N$, study the large order behavior of the series (\ref{fNSW}), and see whether it agrees with the 
prediction of (\ref{z10SW}). In particular, one can test that the correct value of $\Omega(N)$ is indeed the one given in (\ref{omstrong}). 
To do this, we proceed as follows. At leading order in $g_s$, (\ref{z10SW}) gives 
\be
\mathfrak{Z}^{(1|0)}(g_s;N) \approx \re^{2 \pi \ri N} \mu(N) g_s^{1-2N} \re^{8/g_s}, 
\ee
where 
\be
\label{fn-ex}
\mu(N)= -{2^{8N-5} \over \Gamma(N) \Gamma(N-1) }. 
\ee
At the same time, the value of $\mu(N)$ can be extracted numerically from the large order behavior of (\ref{fNSW}). To do this, we assume a large order behavior of the form (\ref{asym-10}), where $\CA=-8$ and $b=2N-2$, and we include as well the contribution of the negative instantons. We determine the leading coefficient by generalized Richardson extrapolation, which eliminates the unknown coefficients of the retained power and logarithmic corrections, including the contribution of the opposite instanton action. 
We compare the predicted value of $\mu(N)$ and the numerical calculation in \figref{fncomp-fig}. The prediction agrees with the 
data up to $15$ digits. One can also make a 
comparison for negative and even complex values of $N$, and consider the subleading corrections to the asymptotics due to higher order terms in $g_s$ in (\ref{z10SW}). We find again excellent agreement. 
Our conclusion is that, indeed, the instanton of the conventional perturbative series ``selects" the 
strong coupling region of the large $N$ theory. In addition, we find no evidence of a wall-crossing 
phenomenon in the large order behavior of the perturbative series as we change $N$.

\sectiono{Example 3: local Calabi--Yau manifolds }
\label{sec-cy}

In this section, we illustrate the general theory with examples of topological strings on local CY manifolds. 
We first consider the resurgent structure of the topological string free energies in the large $N$ theory. Then, 
we will derive some of the results of \cite{gm-peacock} on the resurgent structure of the perturbative 
series, from their large $N$ counterparts. Our examples are the well-known cases of local $\IF_0$ and local $\IP^2$, which 
have been extensively discussed in the literature on local mirror symmetry. 

\subsection{Local $\IF_0$}

The mirror curve to local $\IF_0$ is given by
\begin{equation}
  \re^x + m_{\IF_0} \re^{-x} + \re^y +\re^{-y} + z^{-1/2} = 0,\quad x,y\in\IC.
  \label{eq:mirrorF0}
\end{equation}
The modulus describing the complex structure is $z$, and in addition we have a so-called 
mass parameter $m_{\IF_0}$.  For simplicity we only consider
$m_{\IF_0} = 1$, in which case the conifold point is located
at
\begin{equation}
  z = \frac{1}{16}.
\end{equation}
There are two periods or flat coordinates. The vanishing period at the conifold is
\be
t_c= {2 \over \pi} \int_0^{1-16z} {K(y) \over 1-y} \rd y =  {1\over \pi^2} G_{3,3}^{3,2}\left(16 z\left|
\begin{array}{c}
 \frac{1}{2},\frac{1}{2},1 \\
 0,0,0
\end{array}
\right.\right)-\pi, 
\ee
while the other period gives the complexified K\"ahler parameter  
\be
t(z)=- \log (z) -4 z \, _4F_3\left(1,1,\frac{3}{2},\frac{3}{2};2,2,2;16 z\right), 
\ee
and one has
\be
\label{tconF0}
2 \pi t\left( {1\over 16} \right) = 16 C , 
\ee
where $C$ is Catalan's constant. In this section we denote the conifold period by 
$t_c$, with an explicit subscript, and $t$ will be used for the large radius period, or complexified K\"ahler parameter. 
The genus zero free energy in the conifold frame is given by 
\be
\label{F0F0}
\CF_0(t_c)= {t_c^2 \over 2} \left( \log \left( {t_c \over 16} \right)-{3\over 2} \right)- 8 C t_c-{t_c^3 \over 48}+ \CO(t_c^4). 
\ee
As in the case of Chern--Simons theory on $\IS^3$, the conifold free energies we will consider are normalized as in the matrix model case (\ref{fg-cr}), (\ref{con-free}), (\ref{fg-con}). In particular, we do not include the constant map terms appearing in e.g. \cite{kmz}. Their regularized version thereof is obtained, as usual, by subtracting the conifold part $\CF_g^{\rm con}(t_c)$. 

\begin{figure}
  \centering
  \includegraphics[height=6cm]{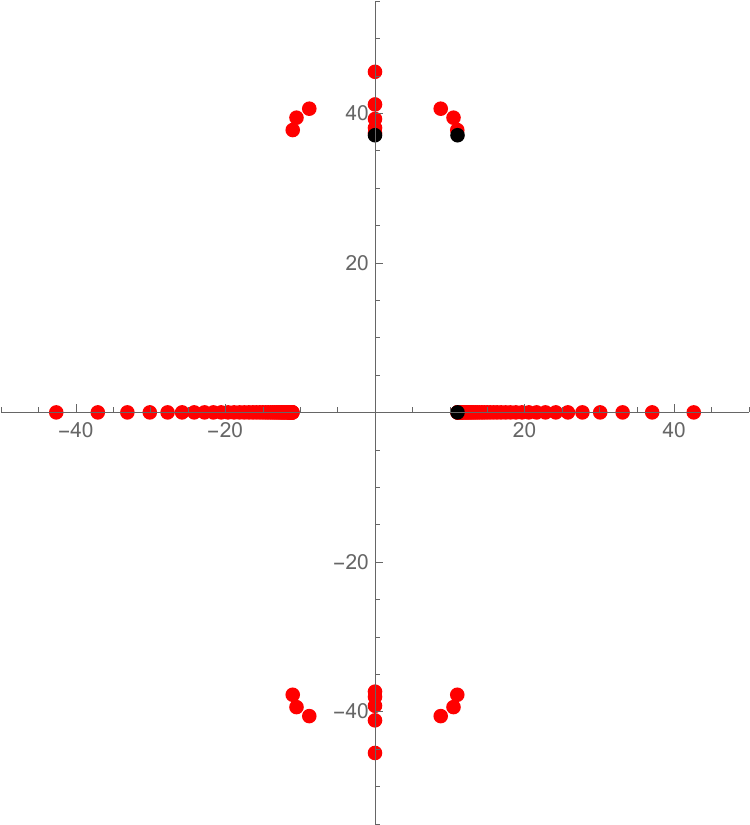}\qquad \qquad  \includegraphics[height=6cm]{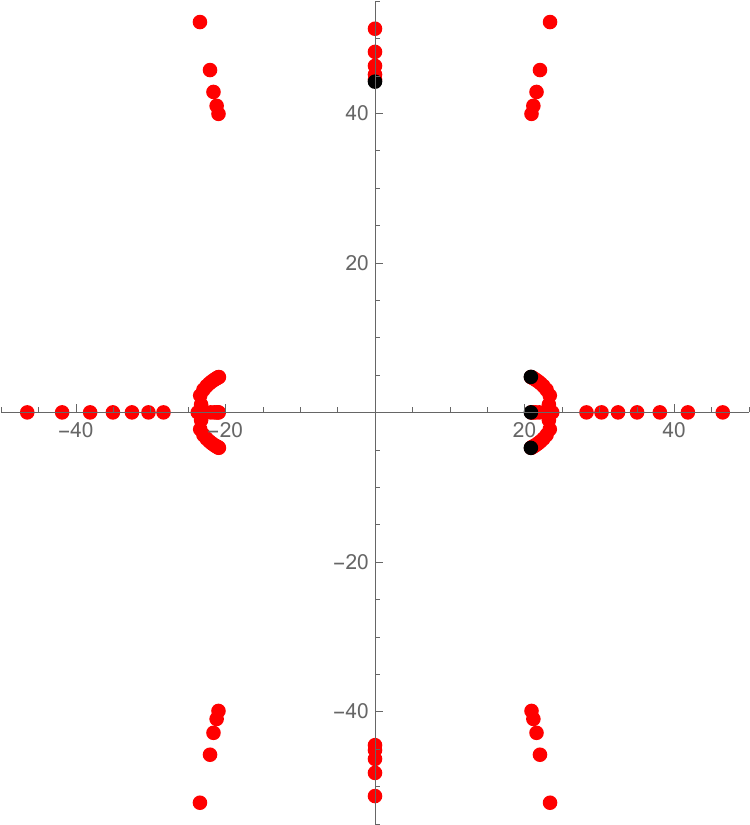} 
  \caption{The Borel singularities of $\CF^{\rm reg} (t_c)$ in local $\IF_0$, for $z=3/32>1/16$ (left), and $z=1/32<1/16$ (right). These regions are the analogues of the strong coupling (respectively, weak coupling) region in SW theory. The black dots in the figure in the left signal the value $2 \pi (t +\ri t_c) $ on the real axis, $4 \pi^2 \ri + 2\pi \ri t_c$ on the imaginary axis, and $2 \pi (t + \ri t_c) +4 \pi^2 \ri +2\pi \ri t_c$ in the first quadrant. The black dots in the figure in the right signal the value $2 \pi t$ on the real axis, $4 \pi^2 \ri + 2\pi \ri t_c$ on the imaginary axis, and $2 \pi t \pm  2 \pi \ri t_c$ above (respectively, below) the real axis. There is also a point $2 \pi t + 4 \pi^2 \ri$ in the first quadrant.}  
  \label{crossF0-fig}
\end{figure}

Let us first discuss the resurgent structure of the regularized conifold free energies, 
near the conifold point $z=1/16$. The first thing 
to note is that the structure of Borel singularities 
changes discontinuously as we move along the real line from $z>1/16$ to $z<1/16$, similarly to 
what happens in SW theory. This is illustrated 
in \figref{crossF0-fig}. On the l.h.s. we show a numerical depiction of the Borel singularities for $z=3/32$, with singularities at 
\be
\CA = \pm 2 \pi (t+ \ri t_c), 
\ee
on the real axis. The Stokes constant for these singularities can be calculated numerically and one finds $\Omega=2$. On the r.h.s., e.g. at $z=1/32$, each of these singularities has split into three, at 
\be
\pm 2 \pi t, \quad \pm 2 \pi \left( t  \pm   \ri t_c \right). 
\ee
The singularity at the real axis has now a Stokes constant given by $\Omega=-4$, while the singularities off the real axis have $\Omega=2$. These values were obtained numerically by studying the large order behavior of the perturbative series, for the singularity on the real axis, and by summing residues of the Pad\'e approximant as described in section \ref{sec-sw}, for the singularities off the real axis. 

In addition to the above, we also have Borel singularities at points of the form (\ref{arn}), 
\be
\label{arn-tc}
\CA_{r,n}= 2 \pi \ri r t_c+ 4 \pi^2 \ri  n. 
\ee
We note that these do not seem to undergo wall-crossing as we move in moduli space. The first singularity in this 
tower, corresponding to $r=n=1$, can be clearly seen on both the l.h.s. and the r.h.s. of \figref{crossF0-fig}. 
The corresponding Stokes constant $\Omega_{r,n}$ with $r=n=1$ can be calculated numerically and one obtains, 
with high numerical precision, 
\be
\label{omega11}
\Omega_{1,1}= 6. 
\ee
As we will see in a moment, this is in agreement with a result found in \cite{gm-peacock}. 

Let us now consider the perturbative series, in which we set $t_c= Ng_s$ in the subtracted regular
free energy $\widehat{\CF}^{\rm reg}(t_c;g_s)$ defined in (\ref{lin-quad}). It reads, to the very first orders,
\begin{align}
\label{pertF0}
\mathfrak{F}(g_s;N) &=- \frac{1}{96}  N \left(2 N^2-1\right)g_s+{   N^2 \over 4608}\left(5 N^2+13 \right)g_s^2+\frac{ N \left(-42 N^4-290 N^2+159
   \right)}{368640}g_s^3\\\nonumber
   &+\frac{  N^2 \left(733 N^4+9080 N^2-6663
  \right)}{44236800}g_s^4+\CO\left(g_s^5\right)
\end{align}
What is the resurgent structure of this perturbative series, as a function of $N$? In view of the large $N$ structure, we expect 
singularities at the points 
\be
\label{simplek}
4 \pi^2 \ri k, \qquad k \in \IZ\backslash\{0\}, 
\ee
induced by the singularities (\ref{arn-tc}), and leading to a trans-series 
of the form (\ref{psN}). In particular, for integer $N$, 
the trans-series simplifies to (\ref{iN}). The singularities (\ref{simplek}) were in fact 
observed in \cite{gm-peacock}, in a detailed study of the resurgent structure of the series 
(\ref{pertF0}). Moreover, by using the TS/ST correspondence and the matrix model for local 
$\IF_0$ obtained in \cite{kmz}, a closed form expression was obtained for the generating series 
of the rational numbers $R_k$ appearing in (\ref{iN}). Let us recall this result. We define the $q$-series
\be
\ba
  g(q) =
  &\sum_{n=0}^\infty
    \frac{(q^{1/2};q)_n^2}{(q;q)_n^2}q^{n/2},\qquad  G(q) =
  &\sum_{n=0}^\infty
    \frac{(q^{1/2};q)_n^2}{(q;q)_n^2}q^{n/2}(1+4n).
\ea
\ee
Then, we have
\be
\ba
\sum_{k \ge 1} R_k q^{k/2}&= \log\left( g(q)G(q) \right)\\
&=6 \sqrt{q}-15 q+60 q^{3/2}-\frac{479 q^2}{2}+\frac{5156 q^{5/2}}{5}-4694
   q^3+\CO\left(q^{7/2}\right). 
\ea
\ee
Since all singularities (\ref{arn-tc}) with different 
values of $r$ get together in the conifold limit, we cannot extract from this result the Stokes constants $\Omega_{r,n}$, but only their linear combinations $\Upsilon_n$, defined in (\ref{Romega}). One finds, from the $q$-series above, 
\be
\sum_{n \ge 1} \Upsilon_n q^n= 6 q-18 q^2 + 58 q^3 -232 q^4 +1030 q^5 -4718 q^6+22138 q^7+ \CO(q^8). 
\ee
These are integer numbers, as expected. We note that the first one agrees with (\ref{omega11}), 
which suggests that for $n=1$ there is only a non-zero value for $\Omega_{r,1}$, namely $\Omega_{1,1}=6$. This can be checked numerically. 

As we have seen, in the large $N$ theory there are also towers of singularities at 
\be
\label{ac-1r}
\CA= \pm 2 \pi t + 2 \pi \ri r t_c+ 4 \pi ^2 \ri n, 
\ee
where $r,n \in \IZ$, and the value of $c$ is the minimal one, $c=2$. The corresponding 
conventional instanton amplitudes can be calculated immediately by using (\ref{f10-final}), (\ref{z10-final}), with the appropriate 
value of $\alpha=16$, which follows from (\ref{F0F0}). One finds
\be
\label{z10F0}
 \ba
 \mathfrak{Z}^{(1|0)}(g_s;N)&= \left( {g_s \over 16} \right) ^{ 2(1-N)} {2 \over \Gamma(N) \Gamma(N-1)}
 \exp \left[ \widehat{\CF}^{\rm reg}((N-2)g_s; g_s) \right]\re^{{16 C\over g_s} - {4 \pi^2 \ri n \over g_s}}\\
 &\quad \times \Biggl\{\Omega_n(N) \Biggl[(N-2) \left(\log\left({g_s \over 16}\right)  + \psi(N-1)-1 \right)\\
 &\qquad + {2\pi^2\ri n-8 C \over g_s}+ g_s \partial_{t_c} \widehat{\CF}^{\rm reg}((N-2) g_s;g_s) \Biggr]
 -{N-2\over 2}\Omega_n'(N) \Biggr\},
\ea
\ee
where we used (\ref{tconF0}). When $N=1$ the zero due to the inverse Gamma function in (\ref{z10F0}) cancels against the pole 
in the digamma function, and one finds the simple expression 
\be
\label{trans-N1}
\mathfrak{Z}^{(1|0)}(g_s;1)= 2\, \Omega_n\, \re^{{16 C\over g_s} - {4 \pi^2 \ri n \over g_s} } \exp \left[ \widehat{\CF}^{\rm reg}(-g_s; g_s) \right], 
\ee
where $\Omega_n$ is defined in (\ref{omega-int}). 
This is exactly the structure of the trans-series found in \cite{gm-peacock}\footnote{To compare the results, one should note that our $g_s$ is $-4 \pi^2$ times the $g_s$ used in \cite{gm-peacock}.}. The Stokes 
constants calculated in \cite{gm-peacock} have to be identified with $2 \Omega_n$. According to the result in \cite{gm-peacock}, one has 
\be
\label{Omega-F0}
\sum_{n \ge 0} q^n \Omega_n= 2- 4 q+ 18 q^2- 96 q^3 + 524 q^4-2876 q^5+ \CO(q^6).  
\ee
We note that the Stokes constants $\Omega_n$ appearing in the perturbative series 
should be induced by the Stokes constants of the large $N$ theory. 
However, as in the case of SW theory, in the limit $t_c= Ng_s$, $g_s \rightarrow 0$ we are 
exactly at the conifold point, and it is not clear if the relevant large $N$ resurgent structure is the one 
at strong coupling  $z>1/16$ or the one at weak coupling $z<1/16$, or whether 
the resurgent structure changes as we change $N$. The first Stokes constant in (\ref{Omega-F0}), 
for $n=0$, agrees with the one obtained in the large $N$ theory at strong coupling, just as in the 
case of SW theory, and this suggests that the other values of $\Omega_n$ are the ones inherited 
from the strong coupling phase for all values of $N$. One can also verify numerically that, as a function of $N$, 
\be
\Omega_0(N)=2 \, \re^{2\pi \ri N},  
\ee
 which indicates that only the strong coupling spectrum is relevant for the perturbative series, and that there is no wall-crossing as a function of $N$, as it happened in SW theory. 

The results above lead to an interpretation of the 
integer invariants obtained in \cite{gm-peacock} in terms of Stokes constants of the large $N$ theory, and therefore, conjecturally, in terms of BPS invariants. On the other hand, the BPS invariants of local Calabi--Yau threefolds can be computed with various techniques \cite{esw,blr1,blr2,blr3,bdlp}. In particular, 
very explicit results for local $\IF_0$ have been found in \cite{longhi-ret}. It would be very interesting to compare in detail the results obtained above, building on \cite{gm-peacock}, with the results obtained with other techniques.  

\subsection{Local $\IP^2$}
\label{replica-P2}

The mirror curve to local $\IP^2$ is given by
\begin{equation}
  \re^x +\re^{y} + \re^{-x-y} + z^{-1/3} = 0,\quad x,y\in\IC.
  \label{eq:mirrorP2}
\end{equation}
The modulus describing the complex structure will be also denoted by $z$, and in our conventions the conifold point is located at
\begin{equation}
\label{p2con}
  z = -\frac{1}{27}.
\end{equation}
Relevant background on topological string theory on local $\IP^2$ can be found in e.g. \cite{gm-multi}, whose 
conventions we will follow (with the only exception that the $t_c$ in that paper is our ${\sqrt{3}} t_c$). The resurgent 
structure of the regularized conifold free energies was studied in some detail in \cite{cms,gm-multi}. Near the conifold point we have singularities of the form (\ref{arn-tc}), and also of the form (\ref{ac-1r}). The complexified K\"ahler parameter, as a function of $z$, is 
given by
\be
\label{tzlocalP2}
t(z)=-\log(z) + 6 z \, _4F_3\left(1,1,\frac{4}{3},\frac{5}{3};2,2,2;-27 z\right).
\ee
It has the following value at the conifold point \cite{rv}
\be
t \left( -{1\over 27} \right)= {9 V \over 2 \pi}  +  \pi \ri,
\ee
where 
\be
V= 2\,  {\rm Im} \, {\rm Li}_2 \left(\re^{ \pi \ri/3} \right)
\ee
and we have chosen a particular branch of the logarithm in (\ref{tzlocalP2}). 

Let us discuss briefly what are the implications of the large $N$ resurgent structure for the perturbative series defined by the subtracted regular free energy $\widehat{\CF}^{\rm reg}(t_c;g_s)$ (as in the case of local $\IF_0$ or Chern--Simons theory on $\IS^3$, we remove constant terms from the higher genus free energies). The perturbative series is given by  
\be
\ba
\mathfrak{F}(g_s;N) & = \frac{ N \left(5-2 N^2\right)}{72 \sqrt{3}}g_s+\frac{ N^2
   \left(N^2-1\right)}{7776}g_s^2+\frac{ N \left(14 N^4-50
   N^2+27\right)}{174960 \sqrt{3}}g_s^3\\\nonumber
   &+\frac{ N^2 \left(-1058 N^4+4245
   N^2-3187\right)}{125971200}g_s^4+\CO\left(g_s^5\right). 
\ea
\ee
The singularities of the form (\ref{ac-1r}) lead to instanton amplitudes of the form (\ref{f10-final}). In this geometry we have $c=3$, and one finds
 \be
 \label{f10P2}
\ba
\mathfrak{Z}^{(1|0)}&= \left( {g_s \over \alpha} \right)^{3 (3-2N)/2} { {\sqrt{2 \pi}} \over \Gamma(N)\Gamma(N-1) \Gamma(N-2)}
 \exp \left[ \widehat{\CF}^{\rm reg}((N-3)g_s; g_s)\right]\re^{-{\CA_0 \over g_s}-{4 \pi^2 \ri n\over g_s}}\\
 &\quad \times \Biggl\{\Omega_n(N) \Biggl[-{1\over 2} + 3 (N-3) \left(\log\left({g_s \over \alpha} \right)  + \psi(N-2)-1 \right)\\
 &\qquad + {\CA_0+4\pi^2\ri n \over g_s}+ 3 g_s \partial_t \widehat{\CF}^{\rm reg}((N-3) g_s;g_s) \Biggr]
 -(N-3)\Omega_n'(N) \Biggr\}.
\ea
\ee
In this formula, 
\be
\label{al-p2}
\alpha= 3^{5/2},
\ee
and 
\be
\label{a0}
\CA_0=-2 \pi t \left( -{1\over 27} \right), 
\ee
where we picked the minus sign in (\ref{ac-1r}). The values (\ref{al-p2}), (\ref{a0}) can be obtained from the results for 
the genus zero free energies in \cite{gm-multi}. In (\ref{f10P2}), we have, according to (\ref{omegaN}),
\be
\Omega_n(N)= \sum_r \Omega_{r,n} (-1)^r \re^{ -2 \pi r \ri N} . 
\ee

Let us now make contact with the results of \cite{gm-peacock} for 
integer values of $N$. When $N=1$, 
the amplitude (\ref{f10P2}) vanishes. When $N=2$ one obtains 
\be
\mathfrak{Z}^{(1|0)}_n= 3 \Omega_n 
 \left( {g_s \over \alpha} \right)^{-3/2} {\sqrt{2 \pi}}\exp \left[ \widehat{\CF}^{\rm reg}(-g_s;g_s) \right] \re^{-{\CA_0 \over g_s}
-{4 \pi^2 \ri n\over g_s}}. 
\ee
After taking into account the different conventions, 
this is exactly the result found in \cite{gm-peacock}, including the precise prefactor. 
The Stokes constants in the perturbative theory, $\Omega_n$, were 
computed in \cite{gm-peacock} for the very first values of $n$, and one has
\be
\label{Omega-P2}
\sum_{n \ge 0} \Omega_n q^n= 1-3 q+ 10 q^2 -29 q^3 + 72 q^4-155 q^5+ \CO(q^6). 
\ee
The interpretation of the $\Omega_n$ appearing in this formula is subject to the same 
caveats that we mentioned in the case of local $\IF_0$, in (\ref{Omega-F0}): at the conifold point (\ref{p2con}) 
we have a wall-crossing phenomenon, and the BPS spectrum changes discontinuously, as in SW theory or local $\IF_0$. Further work is needed to know what is the appropriate spectrum controlling the Stokes constants 
appearing in (\ref{Omega-P2}). 

So far we have discussed the Borel singularities of the perturbative series inherited from instanton actions of the form (\ref{ac-1r}). Instanton actions of the form (\ref{arn-tc}) lead to Borel singularities at the points (\ref{simplek}). For integer $N$, they lead to 
amplitudes of the form (\ref{iN}), and for $N=1$ 
they are the only sources of singularities. The Stokes constants $R_n$ associated to these singularities were 
obtained in \cite{gm-peacock}, and further studied in \cite{rella}. From the results in \cite{rella} one finds that 
the ``traced" Stokes constants $\Upsilon_k$ defined in (\ref{Romega}) are given by 
\be
\Upsilon_k= \begin{cases} 0, & \text{if $k\equiv 0$ mod $3$}, \\
3 , & \text{if $k\equiv 1$ mod $3$},\\
-3, & \text{if $k\equiv 2$ mod $3$}. 
\end{cases}
\ee

Another interesting result that can be obtained from our general formalism concerns the replica limit. 
The corresponding perturbative series (\ref{r-oc}) reads in this case
\be
\mathfrak{R}(g_s)= {5 {\sqrt{3}} \over 216} g_s + {{\sqrt{3}}\over 19 440} g_s^3  -{{\sqrt{3}}\over 1469 664}g_s^5+ {7 {\sqrt{3}}\over 377913600}g_s^7+ \CO(g_s^9).
\ee
The large order behaviour of this series can be obtained from the replica limit 
of the instanton amplitudes, given in (\ref{replica-form}), and can be used to obtain the all-orders 
asymptotics of $f_{g,1}$ for $g \gg 1$. Resumming the latter suggests the following exact formula:
\be
\label{fg1-Rn}
f_{g,1}=4 \pi \Gamma(2g) \sum_{n \ge 1} n R_n (4 \pi^2 \ri n)^{-2g}, \qquad g \ge 2.
\ee
By using e.g. Proposition 4.20 in \cite{rella} one can perform the sum in the r.h.s. explicitly, and (\ref{fg1-Rn}) becomes, 
\be
\label{fg1-closed}
 f_{g,1}=  {\sqrt{3}}(-1)^{g} {B_{2g} B_{2g-1}(2/3) \over (2g)!}, \qquad g \ge 2,  
\ee
where $B_n(x)$ is the Bernoulli polynomial. We have verified the formula (\ref{fg1-closed}) up to $g=45$. 

The formula (\ref{fg1-closed}) is very interesting from the point of view of the relationship between 
topological string amplitudes and BPS states. The conifold behavior of the 
free energies given in (\ref{con-free}) and (\ref{fg-con}) has been interpreted and 
derived as due to a BPS state becoming massless at the conifold point 
\cite{strominger, vafa-fate, gv-conifold}. This state originates in a D4 brane wrapping the vanishing cycle 
at the conifold, whose central charge is proportional to $t_c$. However, it is not clear what is the relationship between the 
regular part of the conifold free energies and the spectrum of BPS states. In principle, the resurgent 
structure of the topological strings gives an answer: conjecturally, the Borel singularities  
have the information on the full BPS spectrum; therefore, a systematic inclusion of these states in the formula 
for the large order behavior of the free energies might lead to an exact formula 
decoding the $\CF_g$s in terms of BPS degeneracies. This is complicated by the fact that 
the trans-series associated to the singularities involve the free energies themselves, 
so such an exact formula would require a ``bootstrap" mechanism, very likely in the form of a 
Riemann--Hilbert problem. However, (\ref{fg1-closed}) is a simple illustration of this general idea: 
the coefficient $f_{g,1}$ appearing in the conifold free energies is explicitly due to ``integrating out" a tower of 
D4-D0 bound states.

\sectiono{Conclusions and open problems}
\label{sec-conclu}

In this paper we have addressed the relationship between large $N$ instantons and what we have called conventional instantons, 
i.e. the instantons associated to the conventional perturbative series at fixed $N$. We have argued that, by starting from the 
large $N$ counterpart, one can obtain expressions for the conventional 
instanton amplitudes which are valid for arbitrary (complex) values of $N$, including the replica limit $N=0$. 
Most of our analysis has focused on the cases of matrix models and topological strings, and we have seen that 
the procedure is not completely straightforward. First of all, a regularization has to be used. In addition, the Stokes constants 
associated to the instanton corrections can not be deduced unambiguously from the large $N$ amplitudes, due to the wall-crossing 
phenomena, which have no counterpart at finite $N$. Understanding this last point in detail remains perhaps the most interesting and practical problem raised by this investigation. 

One of the main motivations of the present paper was to provide a computational tool for the calculation of Stokes 
constants in topological string theory, along the lines initiated by \cite{gm-peacock}. In that paper the large $N$ conifold free energies of the topological string were used to define perturbative series for integer values of $N$. These series can be related to traces of operators by the TS/ST correspondence \cite{ghm,mz}, and this could be exploited to obtain explicit results for the Stokes constants of the conventional instanton amplitudes. In this paper we have shown that the instanton amplitudes obtained in \cite{gm-peacock} can be derived from 
their large $N$ counterparts, and the Stokes constants for the perturbative series turn out to be given by ``traced" versions of these conjectural BPS invariants, in which 
one has to sum over the different singularities which coalesce at the conifold point. In addition, in the case of BPS spectra which 
change discontinuously at the conifold point, these ``traced" versions seem to reflect the spectrum at one of the
phases. Therefore, the analysis of this paper shows that the approach of \cite{gm-peacock} can be used to obtain some analytic results on Stokes constants in topological string theory near the conifold point. Currently this approach has various limitations due to the 
caveats just mentioned, but it is likely that a clever use of the TS/ST correspondence can lead to further results on the resurgent 
structure of the topological string. 

Another motivation of this paper was to obtain a framework to better understand the replica limit $N \rightarrow 0$ in theories 
with a $1/N$ expansion. In the case of matrix models the results we have obtained provide a rather detailed description of this limit, but the physically more relevant cases of percolation or quenched disorder seem to be more complicated. In the toy models considered in \cite{mckane,bray} it is indeed possible to consider a $1/N$ expansion, 
but order by order in $1/N$ one obtains a series in the 't Hooft parameter which is factorially divergent, 
as in theories with renormalons (see e.g. \cite{dmss}). This makes it more complicated to understand the resurgent structure of the replica limit in these models, and more work is needed.

\section*{Acknowledgements}
We would like to thank Alba Grassi for suggesting that we look at Seiberg--Witten theory, Jie Gu for discussions and collaboration 
in other projects, Pietro Longhi for communicating to us his calculations of BPS invariants, and David Sauzin for 
sharing his insights and unpublished notes on parametric resurgence. We also thank Alba Grassi, Jie 
Gu and Pietro Longhi for insightful comments on a preliminary version of this paper. Finally, we would like 
to thank Albrecht Klemm for sharing his notes on Seiberg--Witten theory in 2006. We acknowledge the use of Mathematica for 
most of the symbolic and numerical calculations performed in this paper. For the revised version, we were assisted by ChatGPT for error detecting, text editing, and for the calculations that lead to \figref{fncomp-fig}. This work 
has been supported in part by the ERC-SyG project 
``Recursive and Exact New Quantum Theory" (ReNewQuantum), which 
received funding from the European Research Council (ERC) under the European 
Union's Horizon 2020 research and innovation program, 
grant agreement No. 810573.

\appendix

\sectiono{A simple derivation of the multi-instanton formula}
\label{appendix-a}
As is well-known, in order to define the perturbative topological string partition function one has to specify a frame, 
i.e. a choice of flat coordinates on the moduli space of complex structures. The partition functions in different 
frames are related by a Fourier transform \cite{abk}. It is natural to expect that the full partition function trans-series, 
involving non-perturbative corrections, when computed in different frames, are also related by a Fourier transform. 
As we will now see, this provides a simple derivation of the results of \cite{gm-multi,im} for the 
non-trivial multi-instanton amplitudes (\ref{stokesZ}), (\ref{multi-ex}), starting from the simpler Pasquetti--Schiappa 
instanton amplitudes (\ref{stokes-afr}). 

To be concrete, we will consider the relation between the large radius and the conifold frame for topological strings on local CY threefolds. The corresponding flat coordinates will be denoted by $t$, $t_c$. The relation between the periods is 
\be 
\label{st}
\ba
{\partial \CF_0^c \over \partial t_c}&= -{2 \pi \over \mathfrak{c}} t, \\
t_c&= {\mathfrak{c} \over 2 \pi} {\partial \CF_0 \over \partial t} -{\xi \over 2\pi}, 
\ea
\ee
where $\mathfrak{c}$ is the minimum value of $c$ in (\ref{acr}) and it is an integer number 
characteristic of the CY. One has $\mathfrak{c}=2,3$ for local $\IF_0$ and local $\IP^2$, respectively. In (\ref{st}), 
$\xi$ is a constant, equal to $ 2\pi^2/3$ and $\pi^2$ for local $\IF_0$ and local $\IP^2$, respectively (see 
e.g. \cite{mz,kmz}). The results in \cite{abk} show that the perturbative partition functions are related by the formal Fourier transform
\be 
Z^{(0)}(t_c)= \int Z^{(0)} (t)\,  \re^{-{S (t, t_c) \over g_s^2}} \rd t, 
\ee
where in this case 
\be
S(t, t_c)= {2 \pi t t_c \over \mathfrak{c}} + {\xi \over \mathfrak{c}} t. 
\ee
Let us now consider the Pasquetti--Schiappa multi-instanton amplitude (\ref{stokes-afr}) in the large radius frame, with 
\be
\CA= 2 \pi r t, \qquad r\in \IZ. 
\ee
Then, its Fourier transform is 
\be
\int \, \exp\left\{ \ri a \left( {\rm Li}_2 \left({\cal C} \, \re^{-\CA/g_s} \right)-  {\CA  \over g_s} \log\left(1-{\cal C}  \, \re^{-\CA/g_s}\right) \right) \right\} 
Z^{(0)}(t) \,  \re^{-{S (t, t_c) \over g_s^2}} \rd t, 
 \ee
 which can be immediately evaluated to 
 \be
 \label{final-ref}
 \exp \left\{ \ri a \left(  {\rm Li}_2\left({\cal C}  \, \re^{-c g_s \partial_{t_c} } \right)-  c g_s \partial_{t_c}  \log\left(1-{\cal C}  \, \re^{-c g_s \partial_{t_c}} \right) \right) \right\} Z^{(0)}(t_c), 
\ee
where 
\be
c= - r \mathfrak{c}, 
\ee
so that 
\be
\label{action-tc}
\CA= c {\partial \CF_0^c \over \partial t_c}. 
\ee
The expression (\ref{final-ref}) is precisely the result obtained in \cite{im} for an instanton action of the form (\ref{action-tc}), and as shown in that paper is equivalent to the result (\ref{stokesZ}), (\ref{simple-identity}). Conversely, in view of the above 
argument, one can consider the result of \cite{im} as evidence that the full trans-series for the partition function transforms as the 
perturbative one. This also indicates that the position of Borel singularities and the Stokes constants, which appear naturally in the 
trans-series as exponents and prefactors, do not depend on the frame, see \cite{gu-res} for a different argument in this direction. 

Let us point out that the above derivation gives the holomorphic limit of the multi-instanton amplitudes. To obtain the non-holomorphic version, with the full dependence on the propagators, one has to use the operator formalism of \cite{gm-multi,gkkm}. 

\sectiono{A derivation of the multi-instanton formula for the one-cut matrix model}

As a spinoff of the methods developed in this paper, we show here how to obtain the 
one-cut multi-instantons of \cite{multi-multi} from the 
general multi-instanton formula of \cite{gm-peacock, marmir}, which was an open problem stated in \cite{marmir}. 

The formula (\ref{msw-formula}) for the multi-instantons of the one-cut matrix model should be 
a particular case of the general multi-instanton formulae in \cite{gm-multi,im}, but it looks very different. In particular, 
the prefactors appearing in the general multi-instanton formulae are absent from (\ref{msw-formula}). 
To understand this, one should note following \cite{marmir} that the one-cut limit of the two-cut formula is singular, 
and it has to be taken after an appropriate regularization. We will now derive the formula (\ref{msw-formula}) from the general multi-instanton formula (\ref{stokesZ}), (\ref{simple-identity}), by using a regularization which is very similar to the one that we have proposed to derive the multi-instanton amplitudes at fixed $N$. 

For simplicity, we will consider a generic two-cut matrix model with 't Hooft parameters $t_1$, $t_2$, and we will take $t_2$ to zero, but keeping $t_1+t_2=T$ fixed. We will write the total free energies as $\CF(t;T,g_s)$, where 
$t=t_1$, or as $\CF(t_1, t_2;g_s)$. We note that, as in (\ref{decom}), 
\be
\CF(t_1, t_2;g_s)= \widehat{\CF}(t_1, t_2;g_s) + \CF^{\rm con}(t_2;g_s), 
\ee
and the second term in the r.h.s. is singular in the one-cut limit. 
The exponent appearing in the multi-instanton amplitudes can be written as  
\be
\ba
\CF(t- \ell g_s; g_s)- \CF(t; g_s)&=  \widehat{\CF}(t_1-\ell g_s , t_2+ \ell g_s;g_s)- \widehat{\CF}(t_1, t_2;g_s) + 
\CF^{\rm con}(t_2+ \ell g_s;g_s)\\
&-\CF^{\rm con}(t_2;g_s),  
\ea
\ee
where we have set $c=1$, which is the correct value in eigenvalue tunneling. The first two terms are regular as $t_2 \rightarrow 0$. The two last terms can be regulated as in (\ref{reg-plus}). After 
setting $t_2= g_s x$ and exponentiating we find   
\be
\exp\left( \CF^{\rm con}(t_2+ \ell g_s;g_s)-\CF^{\rm con}(t_2;g_s)\right) \rightarrow {G_2 (-x+1-\ell) \over G_2(-x+1)} (2 \pi)^{\ell/2} g_s^{\ell^2/2 + \ell x } \ri^{-\ell^2}\re^{-\ri \pi \ell x}. 
\ee
In the limit $x \rightarrow 0$ we find
\be
{G_2 (-x+1-\ell) \over G_2(-x+1)} = (-1)^{\ell (\ell+1)\over 2} x^\ell G_2(\ell+1) \left( 1+ \CO(x) \right), 
\ee
i.e. the exponent of the multi-instanton amplitude leads to 
a zero of order $\ell$. There is however also a pole of order $\ell$ in the prefactor 
multiplying the amplitude. This is because this prefactor requires a regularization of the form, 
\be
g_s^k \partial_t^k \CF^{\rm con} (t_2+ \ell g_s; g_s)\rightarrow   (-1)^k \partial^k_x \log G_2(-x+1-\ell ) + \cdots
\ee
where we have taken into account that $\partial_{t_1}= - \partial_{t_2}$ since $t_1+t_2=T$ is fixed. To obtain the 
coefficient of the singular term, we can use (\ref{simple-identity}), in which we set $Z= G_2(-x+1)$. It is easy to see that we have to extract the leading singular behavior as $x \to 0$ of 
\be
\label{im-g}
{1\over G_2(-x+1-\ell)} \left[ \re^{\ri a {\rm Li}_2(\CC)} G_2(-x+1 - \ell-\ri a \log(1-\CC)) \right]_{\CC^\ell},
\ee
where the notation means that we expand the function inside the bracket in $\CC$ and pick the coefficient of $\CC^\ell$. A simple analysis shows that the singular part is $(-\ri a)^\ell/x^\ell+ \cdots$, and we conclude that the one-cut limit of the 
multi-instanton amplitude is
\be
\Omega^\ell {g_s^{\ell^2/2}  \over (2 \pi)^{\ell/2}} G_2(\ell+1) \exp\left[ \widehat{\CF}(t_1-\ell g_s , \ell g_s;g_s)- \widehat{\CF}(t_1;g_s) \right],
\ee
since $\widehat{\CF}(t_1, 0;g_s)=\CF(t_1;g_s)$ is the one-cut free energy. This is precisely the expression (\ref{msw-formula}), including a prefactor involving an appropriate Stokes constant. This shows that the one-cut multi-instanton formula of \cite{multi-multi} is indeed a limiting case of the more general formulae in \cite{gm-peacock, marmir}, when the 't Hooft parameter associated to the shrinking cut goes to zero.

\bibliographystyle{JHEP}

\linespread{0.6}
\bibliography{biblio-par}

\end{document}